\documentclass[prd,nofootinbib,english]{revtex4-2}
\usepackage{amsmath}
\usepackage{amsfonts,color}
\usepackage{amssymb,float}
\usepackage[utf8]{inputenc}

\usepackage{color}
\usepackage[unicode=true,pdfusetitle,
 bookmarks=true,bookmarksnumbered=true,bookmarksopen=true,bookmarksopenlevel=3,
 breaklinks=false,pdfborder={0 0 1},backref=false,colorlinks=true,citecolor=blue]
 {hyperref}
\usepackage{enumitem,accents}

\usepackage{tikz}
\usetikzlibrary{shapes.geometric}
\usetikzlibrary{arrows.meta,arrows}

\usepackage[normalem]{ulem}

\newcommand{\lc}[1]{\accentset{\circ}{#1}}

\definecolor{mygreen}{rgb}{0,0.7,0}

\newcommand{\dd}{{\rm d}}

\begin{document}
	\title{Distinctive Features of Hairy Black Holes in Teleparallel Gauss-Bonnet Gravity}

	\author{Sebastian Bahamonde}
	\email{sbahamondebeltran@gmail.com, bahamonde.s.aa@m.titech.ac.jp}
	\affiliation{Department of Physics, Tokyo Institute of Technology
		1-12-1 Ookayama, Meguro-ku, Tokyo 152-8551, Japan.}

	\author{Daniela D. Doneva}
	\email{daniela.doneva@uni-tuebingen.de}
	\affiliation{Theoretical Astrophysics, Eberhard Karls University of T\"ubingen, T\"ubingen 72076, Germany}
         \affiliation{INRNE - Bulgarian Academy of Sciences, 1784  Sofia, Bulgaria}

	\author{Ludovic Ducobu}
	\email{ludovic.ducobu@umons.ac.be}
	\affiliation{Department of Mathematics and Computer Science, Transilvania university of Brasov, Brasov, Romania}
	\affiliation{Nuclear and Subnuclear Physics, University of Mons, Mons, Belgium}
	
	\author{Christian Pfeifer}
	\email{christian.pfeifer@zarm.uni-bremen.de}
	\affiliation{ZARM, University of Bremen, 28359 Bremen, Germany}
	
	\author{Stoytcho S. Yazadjiev}
	\email{yazad@phys.uni-sofia.bg}
	\affiliation{Theoretical Astrophysics, Eberhard Karls University of T\"ubingen, T\"ubingen 72076, Germany}
	\affiliation{Department of Theoretical Physics, Faculty of Physics, Sofia University, Sofia 1164, Bulgaria}
	\affiliation{Institute of Mathematics and Informatics, Bulgarian Academy of Sciences, Acad. G. Bonchev St. 8, Sofia 1113, Bulgaria}
	
\begin{abstract}
We examine the teleparallel formulation of non-minimally coupled scalar Einstein-Gauss-Bonnet gravity. In the teleparallel formulation, gravity is described by torsion instead of curvature, causing the usual Gauss-Bonnet invariant expressed through curvature to decay into two separate invariants built from torsion. Consequently, the teleparallel formulation permits broader possibilities for non-minimal couplings between spacetime geometry and the scalar field. In our teleparallel theory, there are two different branches of equations in spherical symmetry depending on how one solves the antisymmetric part of the field equations, leading to a real and a complex tetrad. We first show that the real tetrad seems to be incompatible with the regularity of the equations at the event horizon, which is a symptom that scalarized black hole solutions beyond the Riemannian Einstein-Gauss-Bonnet theory might not exist. Therefore, we concentrate our study on the complex tetrad. This leads to the emergence of scalarized black hole solutions, where the torsion acts as the scalar field source.
Extending our previous work, we study monomial non-minimal couplings of degrees one and two, which are intensively studied in conventional, curvature-based, scalar Einstein-Gauss-Bonnet gravity. We discover that the inclusion of torsion can potentially alter the stability of the resulting scalarized black holes. Specifically, our findings indicate that for a quadratic coupling, which is entirely unstable in the pure curvature formulation, the solutions induced by torsion may exhibit stability within certain regions of the parameter space. In a limiting case, we were also able to find black holes with a strong scalar field close to the horizon but with a vanishing scalar charge. 
\end{abstract}
	
\maketitle
\section{Introduction}
By the discovery of the Higgs particle \cite{ATLAS:2012yve,CMS:2012qbp}, and the observation of black hole shadows \cite{EventHorizonTelescope:2019dse} and gravitational waves \cite{LIGOScientific:2016aoc}, it is ensured that both scalar fields and black holes exist in nature. Thus, the assumption that scalarized black holes might exist in the Universe is justified by nowadays knowledge. To really observe scalarized black holes (and to see if they are scalarized by the Higgs or other scalar fields), a solid self-consistent theoretical description, and prediction of their properties and observable signatures, is required.

However, on the basis of General Relativity (GR), the existence of such scalarized black holes is limited according to Israel’s no-hair theorem; see \cite{Misner:1973prb,Chrusciel:2012jk}. From this observation, there are many attempts to evade the rigidity of Israel’s no-hair theorem and construct theories of gravity that allow for scalarized black holes. The other way around, when one modifies general relativity to circumvent its shortcomings -- like the necessity for dark matter and dark energy or the search for quantum gravity \cite{CANTATA:2021ktz,Addazi:2021xuf,Abdalla:2022yfr} -- one finds that several modified theories of gravity naturally posses scalarized black hole solutions. 

The most general theory of a real scalar field non-minimally coupled to a spacetime metric, its Levi-Civita connection, and its curvature, with second-order Euler-Lagrange field equations for the metric and the scalar field in four dimensions is the so-called Horndeski theory \cite{Horndeski:1974}. Among the many sectors of the Horndeski Lagrangian, an interesting one consists of the so-called scalar Gauss-Bonnet (sGB) gravity model, which also emerges from effective field theory from a low-energy string theory limit. In this model, the scalar field is non-minimally coupled to the Gauss-Bonnet invariant of the Levi-Civita connection. In four dimensions the  Gauss-Bonnet invariant is just a total divergence and only through such non-minimal coupling to the scalar field a nontrivial contribution to the field equations is allowed. 

The sGB sector of Horndeski gravity has been extensively studied in the context of black hole physics. In \cite{Kanti:1995vq,Torii:1996yi,Pani:2009wy,Sotiriou:2014pfa}, the authors provide a construction of scalarized black holes in a theory presenting a linear non-minimal coupling term between the scalar field and the Gauss-Bonnet invariant. These were obtained as an explicit counterexample to the no-hair theorem formulated in \cite{Hui:2012qt} for the shift-symmetric sector of Horndeski gravity.
In \cite{Silva:2017uqg,Doneva:2017bvd}, the authors obtain scalarized black hole solutions as a result of spontaneous scalarisation. A quadratic non-minimal coupling function was employed in \cite{Silva:2017uqg} but unfortunately, these scalarised solutions have been proven to be unstable under linear perturbations \cite{Blazquez-Salcedo:2022omw}. Stability can be retained, though, with other coupling functions having a polynomial form \cite{Minamitsuji:2018xde,Macedo:2019sem}.
Even more convenient is to employ an exponential function to define the non-minimal coupling \cite{Doneva:2017bvd}. It allows one to construct hairy black holes via spontaneous scalarization which were further proved to be stable under linear perturbations \cite{Blazquez-Salcedo:2018jnn,Blazquez-Salcedo:2020rhf}. 

The non-minimal scalar-geometry couplings from which Horndeski (and in particular its sGB sector) is constructed assume that the geometrical description of gravity is based on pseudo-Riemannian geometry. However, there are other geometrical descriptions of gravity where torsion (antisymmetric part of the connection) and/or nonmetricity (metric compatibility is absent) are also present, which are known as metric-affine theories of gravity~\cite{Hehl:1994ue,Obukhov:2006gea,Obukhov:2022khx}. Our study would be related to a particular case that is labeled as torsional teleparallel gravity where both curvature and nonmetricity are vanishing and the dynamics of the gravitational interaction is purely encoded into torsion~\cite{Aldrovandi:2013wha,Maluf:2013gaa}. The simplest torsional teleparallel theory produces the same equations as GR with torsion being the field strength tensor of the gravitational field. This theory is known as the Teleparallel Equivalent of General Relativity (TEGR). One can then modify that theory by introducing other torsion invariants or new extra degrees of freedom, such as is done in $f(T)$ gravity, or its extension $f(T,B)$ gravity~\cite{Ferraro:2006jd,Bengochea:2008gz,Bahamonde:2015zma}.

As in the curvature-based theories, we can consider a scalar field non-minimally coupled to torsion~\cite{Ferraro:2006jd,Bengochea:2008gz,Ferraro:2008ey,Geng:2011aj,Bahamonde:2015hza,Hohmann:2018vle,Hohmann:2018dqh,Hohmann:2018ijr,Zubair:2016uhx,Bahamonde:2019shr}. For a comprehensive review of those theories, see~\cite{Bahamonde:2021gfp}. The study of black hole physics in teleparallel gravity has a short history since the majority of the community has been focused on cosmology and also, it has been difficult to obtain asymptotically flat black hole solutions violating the no-hair theorem. One exact solution is the one obtained in~\cite{Bahamonde:2021srr} for a Born-Infeld $f(T)$ gravity whose metric is a generalization of Schwarzschild and its astrophysical properties have been analysed in~\cite{Jusufi:2022loj,Bahamonde:2022jue}. Other solutions, with scalar hair, have also been obtained in scalar-torsion theories with couplings between the torsion scalar and the boundary term~\cite{Bahamonde:2022lvh}.

Recently, it has been found that Teleparallel scalar-torsion theories of gravity providing a non-minimal coupling term extending the Teleparallel Gauss-Bonnet invariant~\cite{Kofinas:2014owa,Bahamonde:2016kba} can support a spontaneous scalarisation mechanism; see \cite{Bahamonde:2022chq}. Interestingly, the spectrum of solutions was shown to naturally extend the one known in the Riemannian theory containing a non-minimal coupling to the Gauss-Bonnet invariant. In that paper, the non-minimal coupling function was chosen to have an exponential form. A question directly following these results is whether those interesting features can be observed for a broader class of coupling functions and whether some qualitative differences can appear with respect to the Riemannian case.

In this paper, we carry on the study started in our previous paper \cite{Bahamonde:2022chq} regarding  Teleparallel Gauss-Bonnet scalarized black hole solutions by investigating the case of linear and quadratic monomial coupling functions. One interest in studying these specific forms for the non-minimal coupling functions comes from the fact that these have attracted considerable attention in the Riemannian case (i.e. in Horndeski gravity); see, for example, Refs.~\cite{Sotiriou:2014pfa} and \cite{Silva:2017uqg}. In addition, a natural comparison with existing results in the Riemannian case can be performed which will enable us to better understand the differences in features and stability between the two formulations.

This paper is organised as follows. In Sec.~\ref{sec:TEGRintro} we briefly review the main aspects of Teleparallel gravity and TEGR necessary for the understanding of the present work. 
Sec.~\ref{sec:sphersym} summarizes and slightly extends the results presented in our previous paper \cite{Bahamonde:2022chq} on how to work in spherical symmetries for Teleparallel gravity and the corresponding field equations.
Sec.~\ref{sec:num} contains the original results of this paper. In Sec.~\ref{sec:numlin}, we investigate the teleparallel extension of the so-called shift-symmetric scalarized black holes where the theory enjoys an extra symmetry under the transformation $\psi \to \psi + \alpha$ for a fixed constant $\alpha$. In Sec.~\ref{sec:numquad}, we consider the case of a quadratic non-minimal coupling function for which spontaneous scalarization is known to happen in the Riemannian case but for which the hairy black hole solutions are unstable. Our study seems to indicate that the stability property of the solutions may change in the presence of purely teleparallel contributions.
We comment on our findings and conclude this discussion in Sec.~\ref{sec:concl}.

In this paper, we work with the metric signature $(+---)$ and we use Latin indices for tangent spacetime indices and Greek indices for spacetime indices. Further, an upper $\circ$ symbol will be used to denote quantities computed with the Levi-Civita connection (Riemannian case).

\section{Introduction to Teleparallel Gauss-Bonnet gravity}\label{sec:TEGRintro}
Teleparallel gravity is an alternative description of gravity based on torsion instead of curvature. In those theories, one considers a metric-compatible connection that has vanishing curvature but possesses torsion. Hence, the gravitational dynamics are only associated with torsion. For its mathematical description, it is useful to introduce tetrads $e^a = e^a{}_\mu \dd x^\mu$ as dynamical variables. They are orthonormal coframes of the metric. The torsion tensor then can be written as
\begin{align}
	T^a{}_{\mu\nu} = 2 \left(\partial_{[\mu}e^a{}_{\nu]} + \omega^a{}_{b[\mu} e^b{}_{\nu]}\right)\,,
\end{align}
with $\omega^a{}_{b\mu}$ being the Teleparallel spin-connection that is purely gauge and, then, one can always choose it to be zero (Weitzenb\"ock gauge). Further, the metric can be reconstructed from the tetrads by taking
\begin{eqnarray}\label{metric}
    g_{\mu\nu} = \eta _{ab} e^a{}_\mu e^b{}_{\nu}  \,,\quad \textrm{with} \quad e^a{}_\mu e_b{}^\mu = \delta ^a_b\,,\quad  \textrm{and}\quad  e^a{}_{\mu}e_a{}^\nu = \delta _\mu ^\nu\,,
\end{eqnarray}
where $\eta_{ab}=\textrm{diag}(1,-1,-1,-1)$ is the Minkowski metric, showing that the tetrads locally provide sufficient information to describe the spacetime geometry.

One can then construct different gravitational theories formulated in terms of torsion. The simplest one of them is constructed from the scalar 
\begin{eqnarray}
	T=\frac{1}{4} T_{\alpha\beta\gamma}T^{\alpha\beta\gamma}+\frac{1}{2}T_{\alpha\beta\gamma}T^{\beta\alpha\gamma}-T^{\lambda}{}_{\lambda\mu}T_{\beta}{}^{\beta\mu}\,.\label{scalarT}
\end{eqnarray}
The choice of this invariant as a Lagrangian gives field equations equivalent to the Einstein equations of GR, and thus this theory is known as TEGR. One can then construct modified teleparallel theories by altering the Lagrangian. For a detailed description of this formalism and different theories, see~\cite{Aldrovandi:2013wha,Bahamonde:2021gfp}. 

In our paper, we are interested to study a particular theory within Teleparallel gravity which is based on the so-called Teleparallel Gauss-Bonnet invariants which are related to the Riemannian Gauss-Bonnet invariant as~\cite{Kofinas:2014owa,Bahamonde:2016kba}
\begin{eqnarray}
	\lc{G}= T_{G}+B_G\,,\label{GB}
\end{eqnarray}
where the Riemannian part is defined as
\begin{eqnarray}
	\lc{G}=\lc{R}_{\alpha\beta\mu\nu}\lc{R}^{\alpha\beta\mu\nu}-4\lc{R}_{\alpha\beta}\lc{R}^{\alpha\beta}+\lc{R}^2\,,
\end{eqnarray}
and the Teleparallel Gauss-Bonnet invariants are
\begin{eqnarray}
T_G&=&\delta^{\mu\nu\sigma\lambda}_{\alpha\beta\gamma\epsilon}K^\alpha{}_{\chi\mu}K^{\chi \beta}{}_\nu K^\gamma{}_{\xi \sigma}K^{\xi \epsilon}{}_\lambda+2\delta^{\mu\nu\sigma\lambda}_{\alpha\beta\gamma\epsilon}K^{\alpha\beta}{}_\mu K^\gamma{}_{\chi\nu}K^{\chi\epsilon}{}_{\xi}K^\xi{}_{\sigma\lambda}+2\delta^{\mu\nu\sigma\lambda}_{\alpha\beta\gamma\epsilon} K^{\alpha\beta}{}_\mu K^{\gamma}{}_{\chi\nu} \nabla_\lambda K^{\chi \epsilon}{}_\sigma\,,\label{TG}\\
	B_G&=&\frac{1}{e}\partial_{\mu}
	\Big[e\delta^{\mu\nu\sigma\lambda}_{\alpha\beta\gamma\epsilon}K^{\alpha\beta}{}_{\nu}\Big(  K^{\gamma}{}_{\xi \sigma}K^{\xi \epsilon}{}_\lambda-\frac{1}{2}\lc{R}^{\gamma\epsilon}{}_{\sigma\lambda}\Big)\Big]\,.\label{BG}
\end{eqnarray}
The quantity 
\begin{align}
    K^{\rho}{}_{\mu\nu}  =\Gamma^{\rho}{}_{\mu\nu}-\lc{\Gamma}^{\rho}{}_{\mu\nu}=\frac{1}{2}\left(T_{\mu}{}^{\rho}{}_{\nu}+T_{\nu}{}^{\rho}{}_{\mu}-T^{\rho}{}_{\mu\nu}\right)\,
\end{align} 
is known as the contortion tensor and $e=\textrm{det}(e^a{}_\mu)$. Further, the covariant derivative appearing in the above equation is explicitly 
\begin{align}
    \nabla_\lambda K^{\chi \epsilon}{}_\sigma=\partial_\lambda K^{\chi \epsilon}{}_\sigma+(\mathring{\Gamma}+K)^\chi{}_{\beta\lambda}K^{\beta \epsilon}{}_\sigma+(\mathring{\Gamma}+K)^\epsilon{}_{\beta\lambda}K^{ \chi\beta}{}_\sigma-(\mathring{\Gamma}+K)^\beta{}_{\sigma\lambda}K^{\chi\epsilon}{}_\beta\,.
\end{align}
Clearly, the Riemannian Gauss-Bonnet invariant is split into two pieces in the Teleparallel gravity language: the first quantity $T_G$ is a topological invariant in four dimensions and the second one $B_G$ is a boundary term in all dimensions. 

Since the Teleparallel Gauss-Bonnet invariants are boundary terms in four dimensions, the simplest theory with them will not give any different dynamic. However, following a similar approach as in the Riemannian sector~\cite{Kanti:1995vq,Torii:1996yi,Pani:2009wy,Sotiriou:2014pfa}, one can couple each of them separately with a scalar field as 
\begin{align}
		\mathcal{S}_{\rm TsGB} 
		&=\frac{1}{2\kappa^2}\int \Big[-T-\frac{1}{2}\beta\, \partial_\mu \psi \partial^\mu \psi
		+\alpha_2 \mathcal{G}_2(\psi)B_G+\alpha_1 \mathcal{G}_1(\psi)T_G\Big] \, e\,\dd^4x
	\end{align}
to acquire dynamics and give a different theory than TEGR/GR or sGB. The first term, $-T$, is the Lagrangian of TEGR defined in~\eqref{scalarT}. The second term is just a kinetic term for the scalar field, and the last two are couplings between the scalar field and the Teleparallel Gauss-Bonnet invariants. By using~\eqref{GB} and introducing the coupling function $\alpha_3\mathcal{G}_3(\psi)=\alpha_1 \mathcal{G}_1(\psi) - \alpha_2 \mathcal{G}_2(\psi)$,  we can rewrite the action in a way that looks more similar to its Riemannian analog, giving
\begin{align}
\mathcal{S}_{\rm TsGB} 
		&=\frac{1}{2\kappa^2}\int \Big[\mathring{R}-\frac{1}{2}\beta\, \partial_\mu \psi \partial^\mu \psi
		+\alpha_2 \mathcal{G}_2(\psi)\mathring{G}+\alpha_3\mathcal{G}_3(\psi) T_G\Big] \, e\,\dd^4x
		\,.\label{action2F}
\end{align}
This form nicely demonstrates which kind of new theories beyond sGB become available due to the use of torsion, instead of curvature. In particular, one can study theories that just employ the terms $T_G$ or $B_G$ to generate torsion-induced non-minimal coupling to a scalar field. In this parametrization $\alpha_3=0$ is just the usual sGB theory, while we call $\alpha_2=0$ or $\alpha_3\mathcal{G}_3(\psi)=- \alpha_2 \mathcal{G}_2(\psi)$ ($\alpha_1 = 0$) purely torsional theories since they are defined respectively in terms of couplings to $T_G$ or $B_G$ alone.

By taking variations with respect to the tetrad fields (the dynamical variables of our theory), we obtain the field equation:
\begin{eqnarray}
		0&=&2\lc{G}_{\beta}^{\nu}+\frac{1}{2}\beta \delta^{\nu}_{\beta}\partial_{\rho}\psi\partial^{\rho}\psi-\beta\partial_{\beta}\psi\partial^{\nu}\psi+\frac{1}{e}\alpha_2(g_{\rho\beta}\delta_{\lambda}^{\nu}+g_{\lambda\beta}\delta_{\rho}^{\nu})
		\eta^{\kappa\lambda\alpha\xi}\epsilon^{\rho\gamma\sigma\tau}
		\lc{R}_{\sigma\tau\alpha\xi}
		\lc{\nabla}_{\gamma}\partial_{\kappa}\mathcal{G}_2(\psi) \nonumber\\
		&&+\alpha_3\Big(\frac{1}{e}\partial_{\mu}\Big[\eta_{al}(Y^{b[lh]}-Y^{h[lb]}+Y^{l[bh]})e_{h}{}^{\mu}e_{b}{}^{\nu}\Big]+\frac{1}{e}T_{iab}e_{h}{}^{\nu}(Y^{b[ih]}-Y^{h[ib]}+Y^{i[bh]})\nonumber\\
		&&-2\mathcal{G}_3(\psi)\delta^{mbcd}_{ijkl}e_{d}{}^{\nu}K^{ij} {} _m K^{k} {} _ {eb}\partial_a K^{el} {} _c-\mathcal{G}_3(\psi)T_G e_a{}^\nu\Big)e^a{}_\beta\,,\label{fieldequations}
	\end{eqnarray}
 where $Y^b{}_{ij}$ 
\begin{align}
		Y^{b}{}_{ij}& := e \mathcal {G} _ 3 X^b {} _ {ij} - 
		2 \delta^{cabd} _ {elkj}\partial_\mu \Big (e\, \mathcal {G} _ 3 e_d \
		{}^\mu K^{el} {} _c K^k {} _ {ia}\Big)\,, \label{eq:defXY}
	\end{align}
	and
	\begin{eqnarray}
		X^{a}{}_{ij}&=&K_j {}^e {} _b K^k {} _ {fc} K^{fl} {} _d \delta^{abcd} _ {iekl} +  
		K^e {} _ {ib} K^k {} _ {fc} K^{fl} {} _d \delta^{bacd} _ {ejkl} + 
		K^k {} _ {ec} K^{ef} {} _b K_j {}^l {} _d \delta^{cbad} _ {kfil} + 
		K^f {} _ {ed} K^{el} {} _b K^k {} _ {ic}\delta^{dbca} _ {flkj}\nonumber\\
		&&+2 K^k {} _ {eb} K^{el} {} _f K^f {} _ {cd}\delta^{abcd} _ {ijkl} + 
		2 K^{ke} {} _b K_j {}^l {} _f K^f {} _ {cd}\delta^{bacd} _ {keil} + 
		2 K^{el} {} _f K^k {} _ {ib} K^a {} _ {cd}\delta^{fbcd} _ {elkj} + 
		2 K^{fc} {} _d K^
		k {} _ {eb} K^{el} {} _i \eta_ {mj} \delta^{dbma} _ {fckl} \nonumber\\
		&&+2 K^k {} _ {eb}\delta^{abcd} _ {ijkl}\partial_d K^{el} {} _c + 
		2 K^{ke} {} _b \delta^{bacd} _ {keil}\partial_d K_j {}^l {} _c\,.\label{X}
	\end{eqnarray}
 The field equations are limited to second-order derivatives in both the tetrad (and subsequently, the metric) and the scalar field. This characteristic becomes evident when examining the Riemannian Gauss-Bonnet contribution, as the coupling function $\mathcal{G}_2$ does not introduce derivatives higher than the second order. In the case of the contribution from the Teleparallel invariant $T_G$, which is mediated by the coupling $\mathcal{G}_3$, the highest-order derivative appearing in the field equation involves first derivatives of the contortion tensor $K_{abc}$. Given that $K_{abc}$ encompasses at most first derivatives with respect to the tetrad field, it establishes that the theory, at most, manifests second-order derivatives in all fields. 
 
It is worth mentioning that the antisymmetric part of the above equation is in general non-zero and it actually coincides with the variation of the gravitational action with respect to the Teleparallel spin-connection. Then, assuming to work in the Weitzenb\"ock gauge, it is sufficient to consider only the equations obtained by varying the action with respect to the tetrads. The reason is that all degrees of freedom of the theory will be encoded purely in the tetrad which is constrained by the symmetric and antisymmetric part of the field equations. This is a general feature for all teleparallel theories of gravity.

Further, by taking variations with respect to the scalar field, we find
 \begin{eqnarray}\label{KG}
		\beta \lc{\square}\psi +\alpha_2 \dot{\mathcal{G}}_2(\psi)\lc{G}+\alpha_3 \dot{\mathcal{G}}_3(\psi)T_G=0\,,
	\end{eqnarray}
where we can notice that for some couplings the invariants $T_G$ and $\lc{G}$ can trigger the scalar field to have a hair. It was recently shown in~\cite{Bahamonde:2022chq} that the above theory admits spontaneously scalarized black hole solutions different from the Riemannian Gauss-Bonnet case. There, the coupling function was assumed to have an exponential form with a quadratic leading order expansion with respect to $\psi$. In the following, we will examine further this topic by considering a variety of qualitatively different couplings that turn out to provide new and interesting phenomenology. In addition, we consider in detail the possibility of the existence of scalarized black holes in the case of real tetrads, a problem that has not yet been addressed properly.

\section{Spherical symmetry in Teleparallel Gauss-Bonnet gravity}\label{sec:sphersym}
In this section, we will show the equations in spherical symmetry and analyse their behaviour near the horizon. They will serve as boundary conditions for the numerical analysis in Sec.~\ref{sec:num}.
\subsection{Tetrad and field equations}
As the dynamical variable of our theory is the tetrad field, there are more degrees of freedom encoded there than in the Riemannian sector, where only the metric determines the dynamics of the theories. In teleparallel gravity, the six extra degrees of freedom of the tetrads can be set by solving the six extra antisymmetric field equations of the theory. To say that our setup possesses a certain symmetry, we impose that the metric, torsion tensor, and scalar field have the same symmetries (i.e., that they are all invariant under the action of a given group encoding the symmetry).  In the following, we will be interested in static and spherically symmetric setups.

Since the torsion tensor can be purely constructed from the tetrad fields, the static and spherically symmetric conditions imposed on torsion implies that using an appropriate coordinate system $(t, r, \vartheta, \varphi)$, the tetrad field in the Weitzenb\"ock gauge becomes~\cite{Hohmann:2019nat,Bahamonde:2022chq}
\begin{subequations}
\begin{eqnarray}
  e^0{}_\mu &=& \nu A(r)   \cosh \beta(r) \dd t + \xi  B(r) \sinh \beta(r) \dd r\,,\\
  e^1{}_\mu &=& \nu  A(r) \sinh \beta(r)\sin \vartheta \cos \varphi   \dd t+ \xi  B(r) \cosh \beta(r) \sin \vartheta \cos \varphi   \dd r+ \chi C(r) ( \cos \alpha(r) \cos \vartheta  \cos \varphi - \sin \alpha(r)\sin \varphi )\dd \vartheta\nonumber\\
  && -\chi C(r)\sin \vartheta (  \sin \alpha(r) \cos \vartheta  \cos \varphi +  \cos \alpha(r) \sin \varphi )\dd \varphi\,,\\
   e^2{}_\mu &=&\nu  A(r)\sinh \beta (r) \sin \vartheta \sin \varphi \dd t+ \xi  B(r)\cosh \beta(r) \sin \vartheta \sin \varphi  \dd r +\chi  C(r)( \cos \alpha(r) \cos \vartheta  \sin \varphi +   \sin \alpha(r) \cos \varphi ) \dd \vartheta\nonumber\\
   &&+ \chi C(r)\sin \vartheta (\cos \alpha(r)  \cos \varphi - \sin \alpha(r) \cos \vartheta  \sin \varphi )\dd \varphi\,,\\
 e^3{}_\mu &=&    \nu  A(r) \sinh \beta(r)\cos \vartheta  \dd t+ \xi  B(r) \cosh \beta(r)\cos \vartheta  \dd r -\chi C(r)\cos \alpha(r) \sin \vartheta   \dd \vartheta + \chi C(r) \sin \alpha(r) \sin ^2\vartheta   \dd \varphi\,,
\end{eqnarray}
\end{subequations}
where the constants $\{\nu,\xi,\chi\}=\pm 1$. The metric then takes the following general form
\begin{equation}
		\dd s^2=A(r)^2 \,\dd t^2-B(r)^2\, \dd r^2-C(r)^2(\dd\vartheta^2+\sin^2\vartheta \dd\varphi^2)\,.\label{metric}
\end{equation}
 Notice that the functions $\alpha(r)$ and $\beta(r)$ are purely tetrad degrees of freedom since they do not appear in the metric.  In~\cite{Bahamonde:2022chq}, it was shown that the antisymmetric field equations of~\eqref{fieldequations} can be solved in two different ways for $\alpha(r)$ and $\beta(r)$ giving us two different branches of field equations for the Teleparallel Gauss-Bonnet gravity. The first one is a real tetrad given by
 \begin{subequations}\label{tetradreal}
 \begin{eqnarray}
     e_{(1)}{}^0{}_\mu &=&A(r)\dd t\,,\\
        e_{(1)}{}^1{}_\mu &=&\xi   B(r) \sin\vartheta \cos \varphi \dd r + \chi C(r)  \cos \vartheta   \cos \varphi \dd \vartheta -\chi C(r) \sin \vartheta   \sin \varphi \dd\varphi\,,\\
           e_{(1)}{}^2{}_\mu &=&\xi   B(r) \sin \vartheta  \sin \varphi\dd r  +\chi C(r) \cos \vartheta  \sin \varphi \dd\vartheta + \chi  C(r) \sin \vartheta  \cos \varphi \dd\varphi\,,\\
              e_{(1)}{}^3{}_\mu &=&\xi   B(r) \cos \vartheta \dd r  -\chi  C(r) \sin \vartheta \dd \vartheta  \,,
 \end{eqnarray}
 \end{subequations}
while the second one is complex:
\begin{subequations}\label{tetrad2}
\begin{eqnarray}
     e_{(2)}{}^0{}_\mu &=&i B(r)\dd r\,,\\
        e_{(2)}{}^1{}_\mu &=&i   A(r) \sin \vartheta  \cos \varphi \dd t  - C(r) \sin \varphi \dd \vartheta -   C(r)\sin \vartheta  \cos \vartheta  \cos \varphi \dd \varphi\,,\\
           e_{(2)}{}^2{}_\mu &=&i   A(r)\sin \vartheta  \sin \varphi  \dd t+   C(r) \cos \varphi \dd\vartheta  -  C(r)  \sin \vartheta  \cos \vartheta  \sin \varphi \dd\varphi\,,\\
              e_{(2)}{}^3{}_\mu &=&i   A(r) \cos \vartheta \dd t+   C(r)  \sin ^2\vartheta \dd\varphi  \,.
 \end{eqnarray}
 \end{subequations}
 As shown in~\cite{Bahamonde:2022chq}, the form of the scalars appearing in the action~\eqref{action2F} in spherical symmetry for the real tetrad does not depend on the sign parameter $\nu$. That is why, for simplicity, it was set to 1 above. The first tetrad was not fully analysed in~\cite{Bahamonde:2022chq} while the second one was used to find scalarized black hole solutions with an exponential coupling function. Then, there are two sets of field equations depending on the tetrad. One can further take $C(r)=r$ without losing generality.

The field equations~\eqref{fieldequations} and scalar field equation~\eqref{KG} for the first tetrad~\eqref{tetradreal} are
 \begin{subequations}\label{fieldeqreal}
		\begin{eqnarray}
			E^t{}_t :\ 0&=&\frac{ 4B'}{r B^3}-\frac{2}{B^2r^2}+\frac{2}{r^2}-\frac{1}{2B^2} \beta   \psi '^2
			-\frac{(\alpha_3\dot{\mathcal{G}}_3+\alpha_2 \dot{\mathcal{G}}_2) \left(8 \left(B^2-3\right) B' \psi'-8 B \left(B^2-1\right) \psi''\right)}{r^2 B^5}\nonumber\\
			&&+\frac{8 \left(B^2-1\right) \psi'^2 (\alpha_3\ddot{\mathcal{G}}_3+\alpha_2 \ddot{\mathcal{G}}_2)}{r^2 B^4}-\frac{16 \psi'^2 (B-\xi  \chi ) \alpha_3\ddot{\mathcal{G}}_3}{r^2 B^3}+\frac{16 \alpha_3\dot{\mathcal{G}}_3 \left(B' \psi' (B-2 \xi  \chi )+B \psi'' (\xi  \chi -B)\right)}{r^2 B^4}\,,\label{eq:tt}\nonumber \\ \\
			E^r{}_r :\ 0&=&-\frac{4 A'}{r A B^2}-\frac{2}{r^2 B^2}+\frac{2}{r^2}+\frac{\beta }{2 B^2}  \psi '^2   
			+\frac{8 \left(B^2-3\right) A' \psi' (\alpha_3\dot{\mathcal{G}}_3+\alpha_2 \dot{\mathcal{G}}_2)}{r^2 A B^4}\nonumber\\
			&&-\frac{16 A' \psi' (B-2 \xi  \chi ) \alpha_3\dot{\mathcal{G}}_3}{r^2 A B^3}\,,\label{eq:rr} \\
			E^\theta{}_\theta :\ 0&=&-\frac{2A''}{A B^2}+\frac{2A' B'}{A B^3}-\frac{2A'}{r A B^2}+\frac{2B'}{r B^3}-\frac{\beta }{2 B^2}\psi '^2
			-\frac{8 A' \psi'^2 (\alpha_3\ddot{\mathcal{G}}_3+\alpha_2 \ddot{\mathcal{G}}_2)}{r A B^4}+\frac{8 \xi  \chi  A' \psi'^2 \alpha_3\ddot{\mathcal{G}}_3}{r A B^3}\nonumber\\
			&&+\frac{8 \xi  \chi  \alpha_3\dot{\mathcal{G}}_3\left(B A'' \psi'+A' \left(B \psi''-2 B' \psi'\right)\right)}{r A B^4}-\frac{8 (\alpha_3\dot{\mathcal{G}}_3+\alpha_2 \dot{\mathcal{G}}_2) \left(B A'' \psi'+A' \left(B \psi''-3 B' \psi'\right)\right)}{r A B^5}\,,\\
   E_\psi :\ 0&=&
		\beta  \left(\frac{\psi ' \left(r B A'+A \left(2 B-r B'\right)\right)}{r A B^3}+\frac{\psi ''}{B^2}\right) - \frac{(\alpha_3\dot{\mathcal{G}}_3+\alpha_2 \dot{\mathcal{G}}_2) \left(8 \left(B^2-3\right) A' B'-8 B \left(B^2-1\right) A''\right)}{r^2 A B^5}\nonumber\\
		&&+\frac{16 \alpha_3\dot{\mathcal{G}}_3 \left(B A'' (\xi  \chi -B)+A' B' (B-2 \xi  \chi )\right)}{r^2 A B^4}\,.\label{eq:psi}
			\label{eq:thetatheta}
		\end{eqnarray}
	\end{subequations}
  Here, primes are derivatives with respect to the radial coordinate whereas dots are derivatives with respect to the scalar field.

On the other hand, the equations for the complex tetrad~\eqref{tetrad2} are
\begin{subequations}\label{fieldeqsym}
		\begin{eqnarray}
			E^t{}_t :\ 0&=&\frac{4B'}{r B^3}-\frac{2}{B^2r^2}+\frac{2}{r^2}-\frac{1}{2B^2} \beta   \psi '^2	-\frac{(\alpha_3\dot{\mathcal{G}}_3+\alpha_2 \dot{\mathcal{G}}_2) \left(8 \left(B^2-3\right) B' \psi'-8 B \left(B^2-1\right) \psi''\right)}{r^2 B^5}\nonumber\\
			&&+\frac{16 \alpha_3\dot{\mathcal{G}}_3 \left(B' \psi'-B \psi''\right)}{r^2 B^3}+\frac{8 \left(B^2-1\right) \psi'^2  (\alpha_3\ddot{\mathcal{G}}_3+\alpha_2 \ddot{\mathcal{G}}_2)}{r^2 B^4}-\frac{16 \psi'^2 \alpha_3\ddot{\mathcal{G}}_3}{r^2 B^2}\,,\label{eq:ttB}\\
			E^r{}_r :\ 0&=&-\frac{4 A'}{r A B^2}-\frac{2}{r^2 B^2}+\frac{2}{r^2}+\frac{\beta }{2 B^2}  \psi '^2+\frac{8 \left(B^2-3\right) A' \psi' (\alpha_3\dot{\mathcal{G}}_3+\alpha_2 \dot{\mathcal{G}}_2)}{r^2 A B^4}-\frac{16 A' \psi' \alpha_3\dot{\mathcal{G}}_3}{r^2 A B^2}
			\label{eq:rrB}\\
			E^\theta{}_\theta :\ 0&=&-\frac{2A''}{A B^2}+\frac{2A' B'}{A B^3}-\frac{2A'}{r A B^2}+\frac{2B'}{r B^3}-\frac{\beta }{2 B^2}\psi '^2-\frac{8 A' \psi'^2 (\alpha_3\ddot{\mathcal{G}}_3+\alpha_2 \ddot{\mathcal{G}}_2)}{r A B^4}\nonumber\\
			&&-\frac{8 (\alpha_3\dot{\mathcal{G}}_3+\alpha_2 \dot{\mathcal{G}}_2) \left(B A'' \psi'+A' \left(B \psi''-3 B' \psi'\right)\right)}{r A B^5}\,,
			\label{eq:thetathetaB}\\
   	E_\psi :\ 0&=&		\beta  \left(\frac{\psi ' \left(r B A'+A \left(2 B-r B'\right)\right)}{r A B^3}+\frac{\psi ''}{B^2}\right)- \frac{(\alpha_3\dot{\mathcal{G}}_3+\alpha_2 \dot{\mathcal{G}}_2) \left(8 \left(B^2-3\right) A' B'-8 B \left(B^2-1\right) A''\right)}{r^2 A B^5} \nonumber\\
		&&+\frac{16 \alpha_3\dot{\mathcal{G}}_3 \left(A' B'-B A''\right)}{r^2 A B^3}\,.\label{eq:psiB}
		\end{eqnarray}
	\end{subequations}
In the following sections, we will analyse both sets of field equations.

 \subsection{Expansions around horizon}\label{sec:5}
Let us consider the expansion of the above equations near the horizon $r_H$. This can be done by taking
\begin{eqnarray}\label{expanhorz}
		A(r)^2&=&a_1(r-r_H)+a_2(r-r_H)^2+\dots\,,\\
		B(r)^{-2}&=&b_1(r-r_H)+b_2(r-r_H)^2+\dots\,,\\
		\psi(r)&=& \psi_H+\psi_H'(r-r_H)+\psi_H''(r-r_H)^2+\dots\,.
	\end{eqnarray}
By assuming these types of expansions, we ensure that $\textrm{det}(g_{\mu\nu})$ is finite at the horizon as long as $b_1$ is non-vanishing. 

The expansion near the horizon for the complex tetrad was computed in~\cite{Bahamonde:2022chq}. To ensure the regularity of the solution at the horizon, the value of $\psi'_H$ is constrained. Depending on the coupling functions, two branches are possible. The first branch (or boundary condition) is when $\alpha_3\dot{\mathcal{G}}_3(\psi_H)\neq\alpha_2 \dot{\mathcal{G}}_2(\psi_H)$, where the scalar field first derivative at the horizon is (denoted as BC1 in the following sections)
	\begin{eqnarray} \label{eq:BC1}
		\psi'_{H}&=& \frac{r_H }{4 (\alpha_2 \dot{\mathcal{G}}_2-\alpha_3 \dot{\mathcal{G}}_3)}\left(1\pm\frac{1}{\beta}\Big[\beta ^2+\frac{32 (\alpha_3 \dot{\mathcal{G}}_3-\alpha_2 \dot{\mathcal{G}}_2) }{r_H^8}\left\{32 \alpha_3^2 \dot{\mathcal{G}}_3^2 (\alpha_3 \dot{\mathcal{G}}_3-\alpha_2 \dot{\mathcal{G}}_2)+\beta  r_H^4 (3 \alpha_2 \dot{\mathcal{G}}_2+\alpha_3 \dot{\mathcal{G}}_3)\right\} \Big]^{1/2}\right)\nonumber \\
		&&-\frac{8 \alpha_3 \dot{\mathcal{G}}_3}{\beta  r_H^3}\,,\label{rhprime}
	\end{eqnarray}
 while the second branch (or boundary condition) is obtained in the case $\alpha_3\dot{\mathcal{G}}_3(\psi_H)=\alpha_2 \dot{\mathcal{G}}_2(\psi_H)$, where the scalar field first derivative evaluated at the horizon becomes (denoted as BC2 in the following sections)
\begin{eqnarray} \label{eq:BC2}
	\psi'_{H}=
	\displaystyle\frac{8 \alpha_2\dot{\mathcal{G}}_2(\psi_H)}{\beta  r_H^3}\,.
\end{eqnarray}

For the real tetrad, the situation is more complicated. To ensure regularity of the real tetrad field equations~\eqref{fieldeqreal} at the event horizon, one finds that the condition $\dot{\mathcal{G}}_{3}(\psi_H)=0$ must be satisfied at first order in the expansions. That condition implies that at first order a theory having non-trivial well-behaved expansion near the horizon must be indistinguishable from the Riemannian Gauss-Bonnet one. Further, even going up to fourth order in expansions, one systematically obtains that all derivatives of the new coupling function must vanish at the scalar field evaluated at the horizon ($\dot{\mathcal{G}}_{3}(\psi_H)=\ddot{\mathcal{G}}_{3}(\psi_H)=\dddot{\mathcal{G}}_{3}(\psi_H)=\ddddot{\mathcal{G}}_{3}(\psi_H)=0$). In our analysis we have gone up to fourth order since higher orders are computationally much more expensive. Our experience suggests, though, that this pattern will most probably remain the same for higher order. Assuming that this conjecture is true and restricting ourselves to coupling functions $\mathcal{G}_3$ that depends only on $\psi$ in an explicit algebraic form, we can conclude that  $\mathcal{G}_3$ is a constant and thus the theory coincides with the  Riemannian Gauss-Bonnet case. 

Even if the above pattern breaks and a derivative of $\mathcal{G}_{3}(\psi)$ higher than four is nonzero, then $\mathcal{G}_{3}(\psi)$ should contain at least fifth order of $\psi$. This is clearly a realm where even  the  Riemannian Gauss-Bonnet case was not investigated and looks a bit like fine-tuning the coupling. For that reason, we will concentrate only on the complex tetrads in the following sections. 

Before going further, let us also stress that in the following, when working with BC1, we will focus on the branch corresponding to the ``$-$" sign in front of the square root in \eqref{eq:BC1}. The reason is that, contrary to the branch with a ``$+$" sign, this branch smoothly connects to GR's solutions in the limit of vanishing non-minimal couplings.

\section{Scalarization in Teleparallel Gauss-Bonnet gravity with different coupling functions}\label{sec:num}
In \cite{Bahamonde:2022chq}, we focused on one specific form for $\mathcal{G}_2$ and $\mathcal{G}_3$: an exponential function with a leading order expansion being quadratic in term of $\psi$. This choice led, at least in the Riemannian Gauss-Bonnet case, to well-behaved stable, spontaneously scalarized black hole solutions. Our analysis showed that actually either one of the $T_G$ or $B_G$, which form the teleparallel Gauss-Bonnet term, suffices to obtain spontaneously scalarized black hole solutions.

In this section, for simplicity, we will consider the coupling functions $\mathcal{G}_2 = \mathcal{G}_3 = \mathcal{G}$ to be equal. Their relative strength can be adjusted by the choices of the parameters $\alpha_2$ and $\alpha_3$. We explore two popular coupling functions, different from the original studies in \cite{Bahamonde:2022chq}, namely:
\begin{itemize}
    \item $\mathcal{G} = \psi$ where the GR black holes are not solutions of the field equations and the black holes are always endowed with scalar hair, see Sec. \ref{sec:numlin};
    \item $\mathcal{G} = \psi^2$, which leads to black hole scalarization, i.e., the Schwarzschild black hole is always a solution of the field equations, but for small black hole masses, it becomes unstable, giving rise to a spontaneously scalarized branch of solutions. Even though simpler compared to the exponential coupling considered in  \cite{Bahamonde:2022chq}, this choice leads to unstable black hole solutions in the Riemannian Gauss-Bonnet case. Interestingly, this observation might change for a strong enough torsional contribution as detailed below, see Sec. \ref{sec:numquad}.
\end{itemize}
Clearly, in the general case, both couplings above have their first derivative nonzero at the horizon. Thus, according to the studies in Sec. \ref{sec:5}, they cannot lead to regular black holes in the case of real tetrads. For that reason, we will consider in the present section only the case of complex tetrads, similar to \cite{Bahamonde:2022chq}.

\subsection{Shift symmetric linear coupling}\label{sec:numlin}
We will first consider a shift symmetric teleparallel Gauss-Bonnet theory with couplings
\begin{align} \label{eq:LinCoupling}
   \mathcal{G}_2=\mathcal{G}_3=\psi.
\end{align}
These models are characterised by the fact that $\psi={\rm const}$ is not a solution of the field equations and thus the black holes are always endowed with scalar hair. This behaviour is contrary to the case of spontaneously scalarized teleparallel black holes considered in the next section and also in \cite{Bahamonde:2022chq}, which emerged from a coupling that is quadratic in the scalar field. In addition, the field equations are invariant under a simultaneous change of signs of the coupling parameters $\alpha_2$, $\alpha_3$, and the scalar field $\psi$. That is why in the results below we present only certain combinations of signs of $\alpha_2$ and $\alpha_3$, while the opposite signs lead to the same solution for the metric function but with an opposite sign scalar field.  

\begin{figure}
    \includegraphics[scale=0.35]{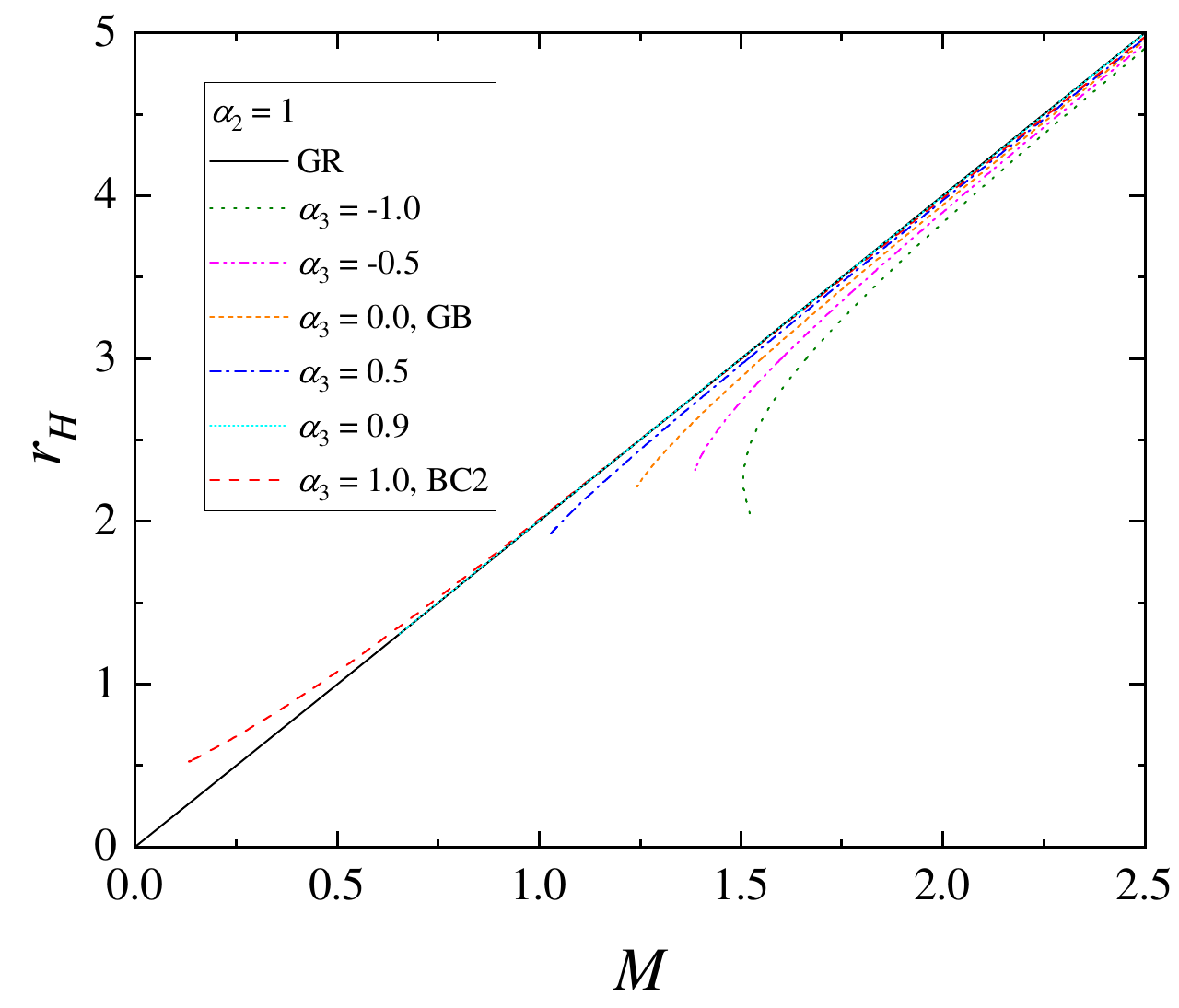}
    \includegraphics[scale=0.35]{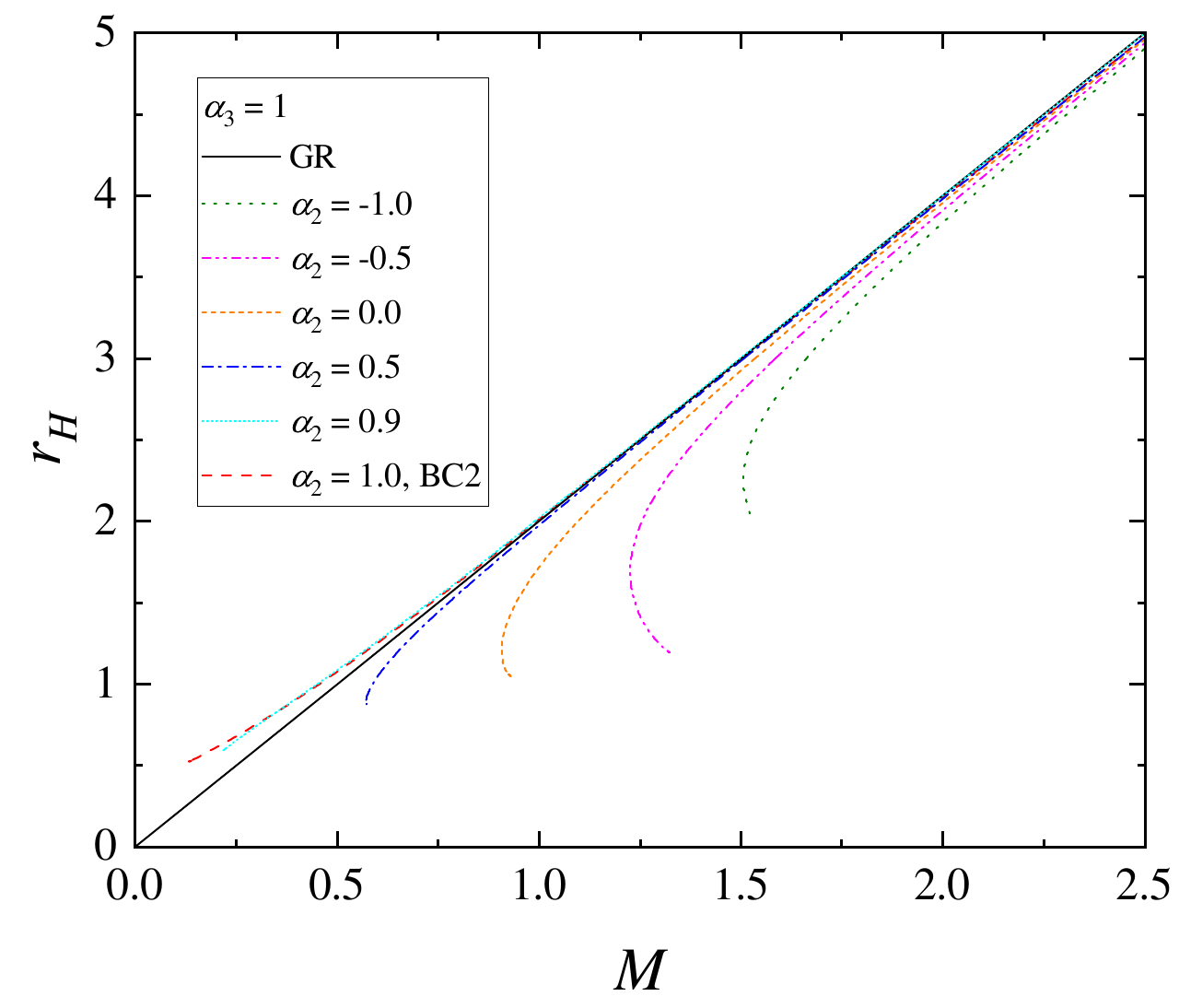}
    \includegraphics[scale=0.35]{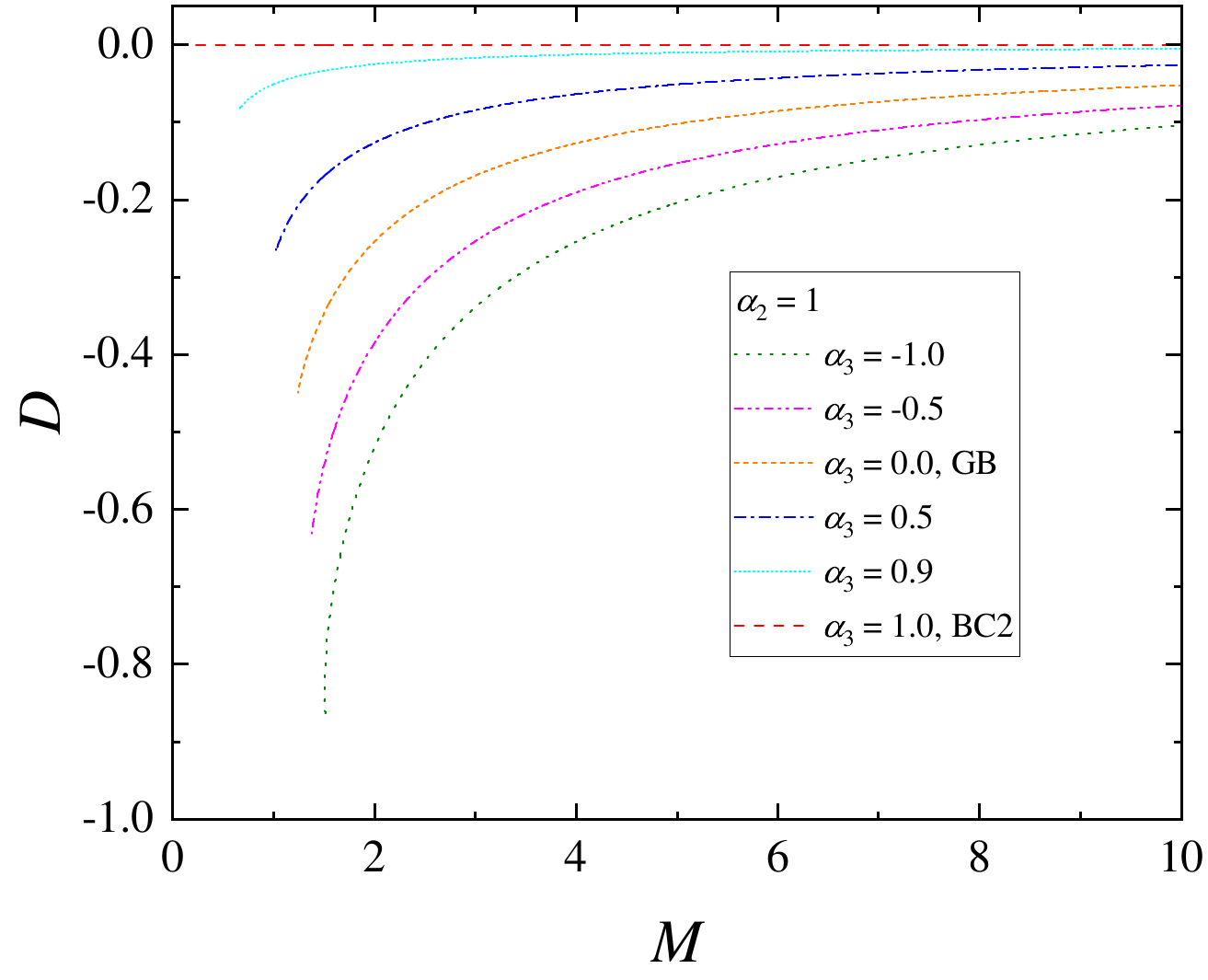}
    \includegraphics[scale=0.35]{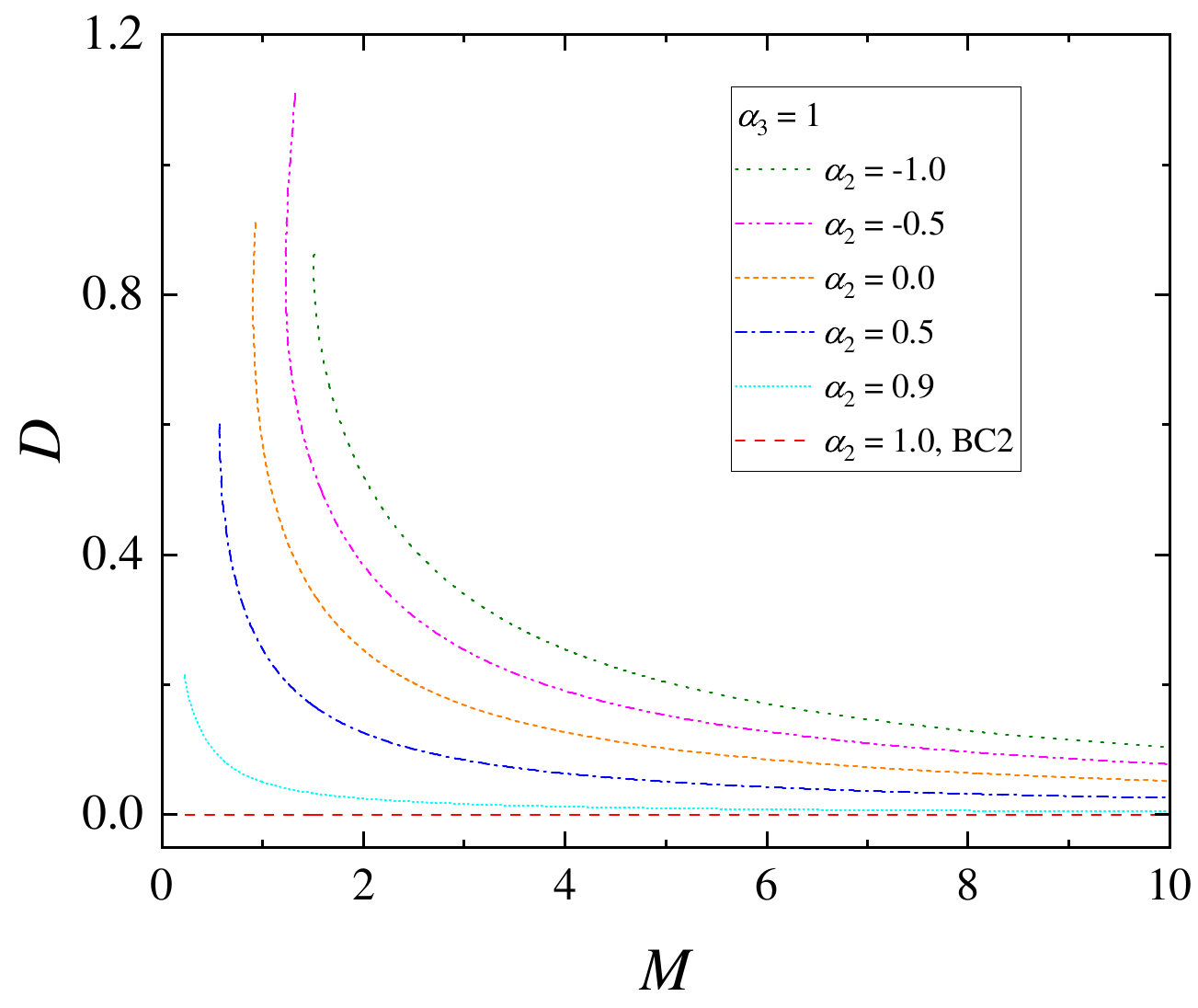}
    \includegraphics[scale=0.35]{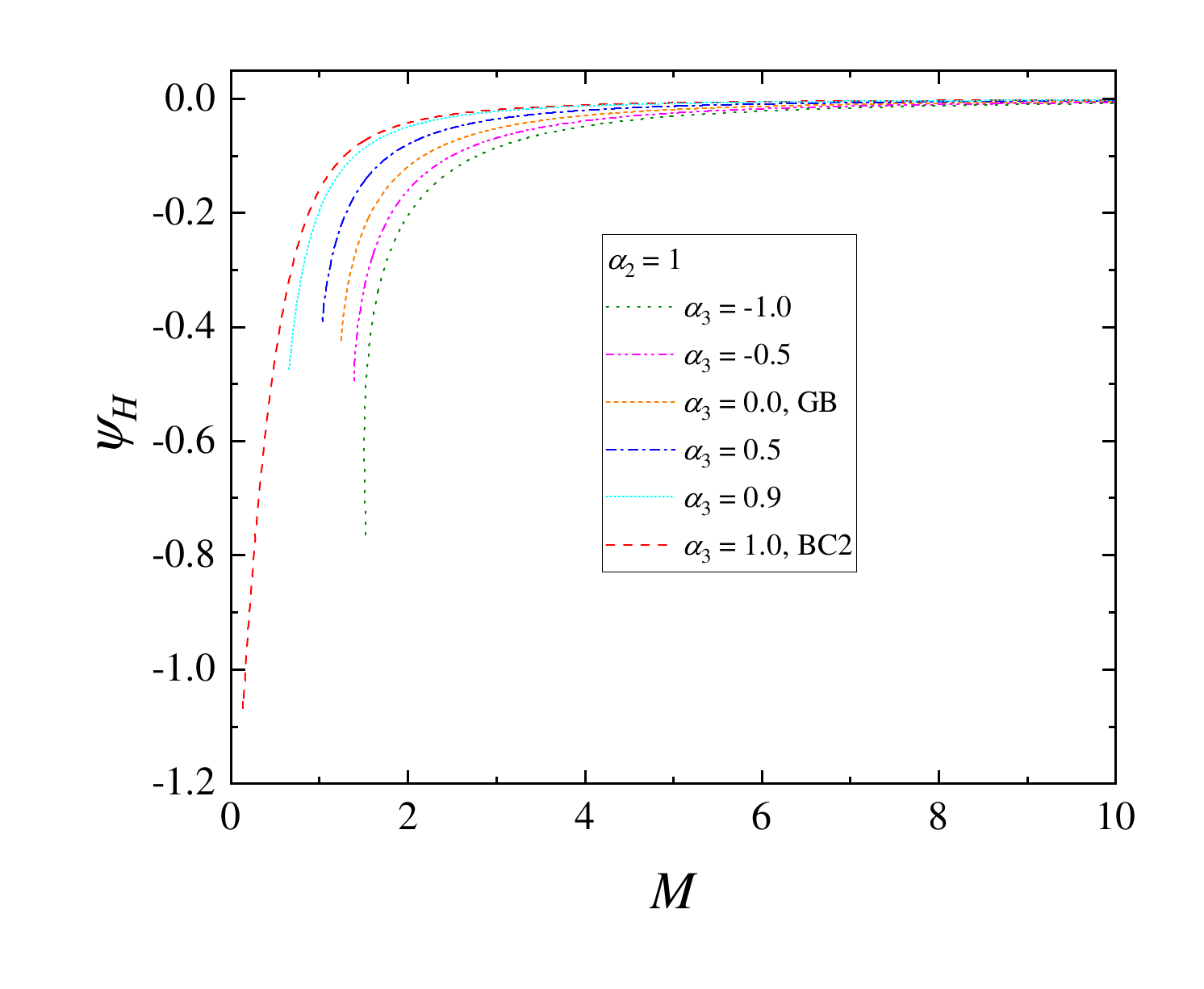}
    \includegraphics[scale=0.35]{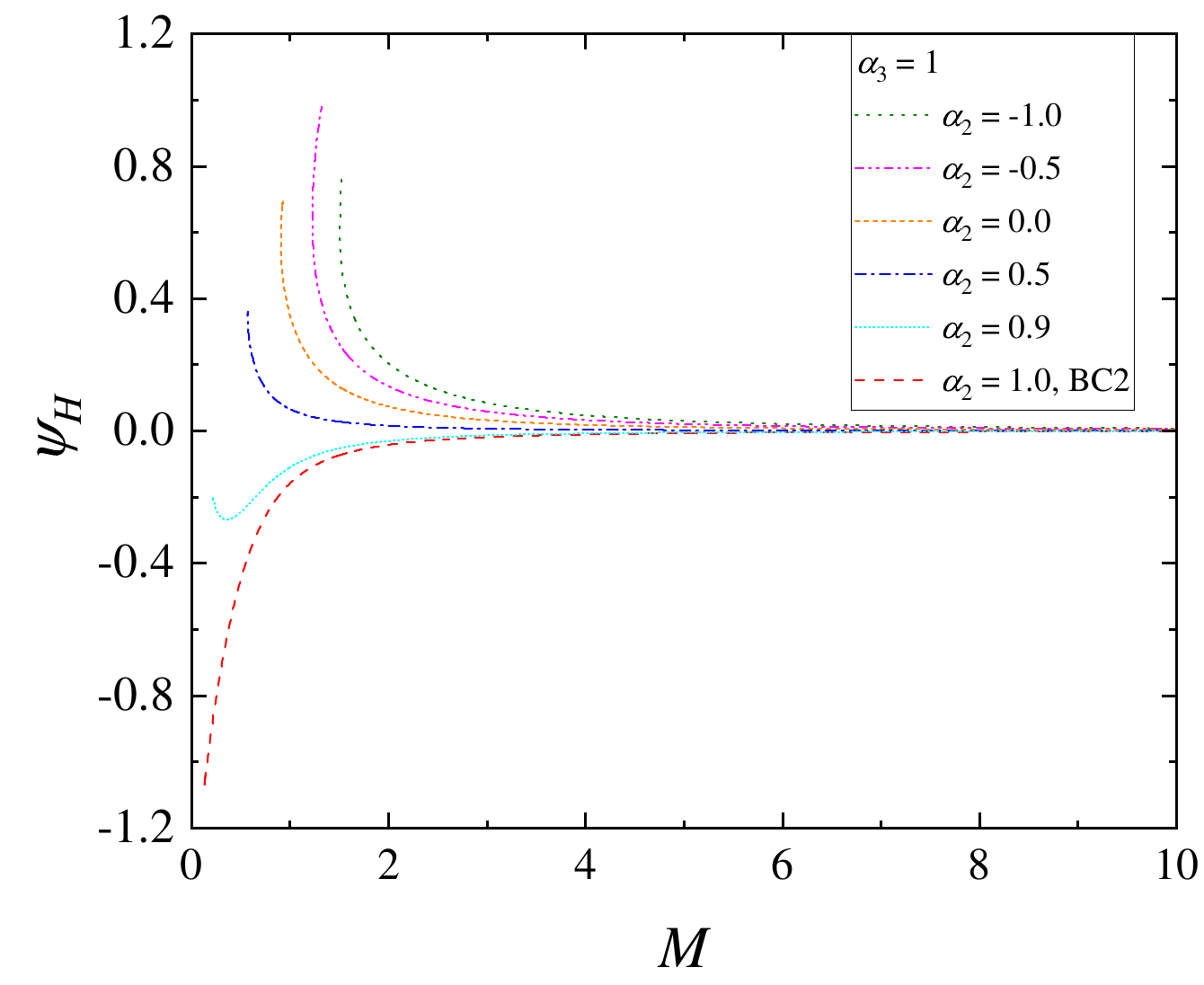}
    \caption{The radius of the horizon (\textit{top panels)}, the scalar charge (\textit{middle panels}), and the scalar field at the horizon (\textit{bottom panels}) as functions of mass for a linear coupling function \eqref{eq:LinCoupling}. The parameters are fixed to $\alpha_2=1$ and varying $\alpha_3$ (\textit{left column}), and $\alpha_3=1$ and varying $\alpha_2$ (\textit{right column}). The bald Schwarzschild black hole is depicted with a solid black line in the top panels. The pure sGB case corresponds to $\alpha_2=-1$ and $\alpha_3=0$.}
    \label{fig:LinCoupling}
\end{figure}

In Fig. \ref{fig:LinCoupling} we present the sequences of black hole solutions for several combinations of parameters. The plots represent the horizon radius, the scalar charge (defined as the leading order asymptotic of the scalar field at infinity $\psi(r\rightarrow\infty)=D/r + \mathcal{O}(1/r^2)$), as well as the scalar field on the horizon as functions of the black hole mass\footnote{The mass we used is the Komar mass of the black hole $\mathcal{M}$ which is extracted from the asymptotic behaviour of $g_{tt}$ and $g_{rr}$ as
\begin{equation*}
\mathcal{M} = \frac{1}{2}\lim_{r\to\infty}\left(r^2 \frac{g'_{tt}}{\sqrt{g_{tt}g_{rr}}}\right)\,.
\end{equation*}}. The two columns of figures correspond to either fixing $\alpha_2=1$ and varying $\alpha_3$ or fixing $\alpha_3=1$ and varying $\alpha_2$. The scalarized branches with $\alpha_3\dot{\mathcal{G}}_3(\psi_H) \ne \alpha_2 \dot{\mathcal{G}}_2(\psi_H)$ employ BC1 (see Eq. \eqref{eq:BC1}) while in the case $\alpha_3\dot{\mathcal{G}}_3(\psi_H) = \alpha_2 \dot{\mathcal{G}}_2(\psi_H)$  BC2 is used (see Eq. \eqref{eq:BC2}). The pure Riemannian sGB theory corresponds to $\alpha_2=-1$ and $\alpha_3=0$. Solutions are typically terminated at the point where the regularity condition at the black hole horizon is violated, namely the expression in the square root of \eqref{eq:BC1} becomes negative. Note that some of the parameter combinations coincide in the left and the right rows which is done for better visualization of the branches' behavior when parameters are varied.

\begin{figure}
    \includegraphics[scale=0.35]{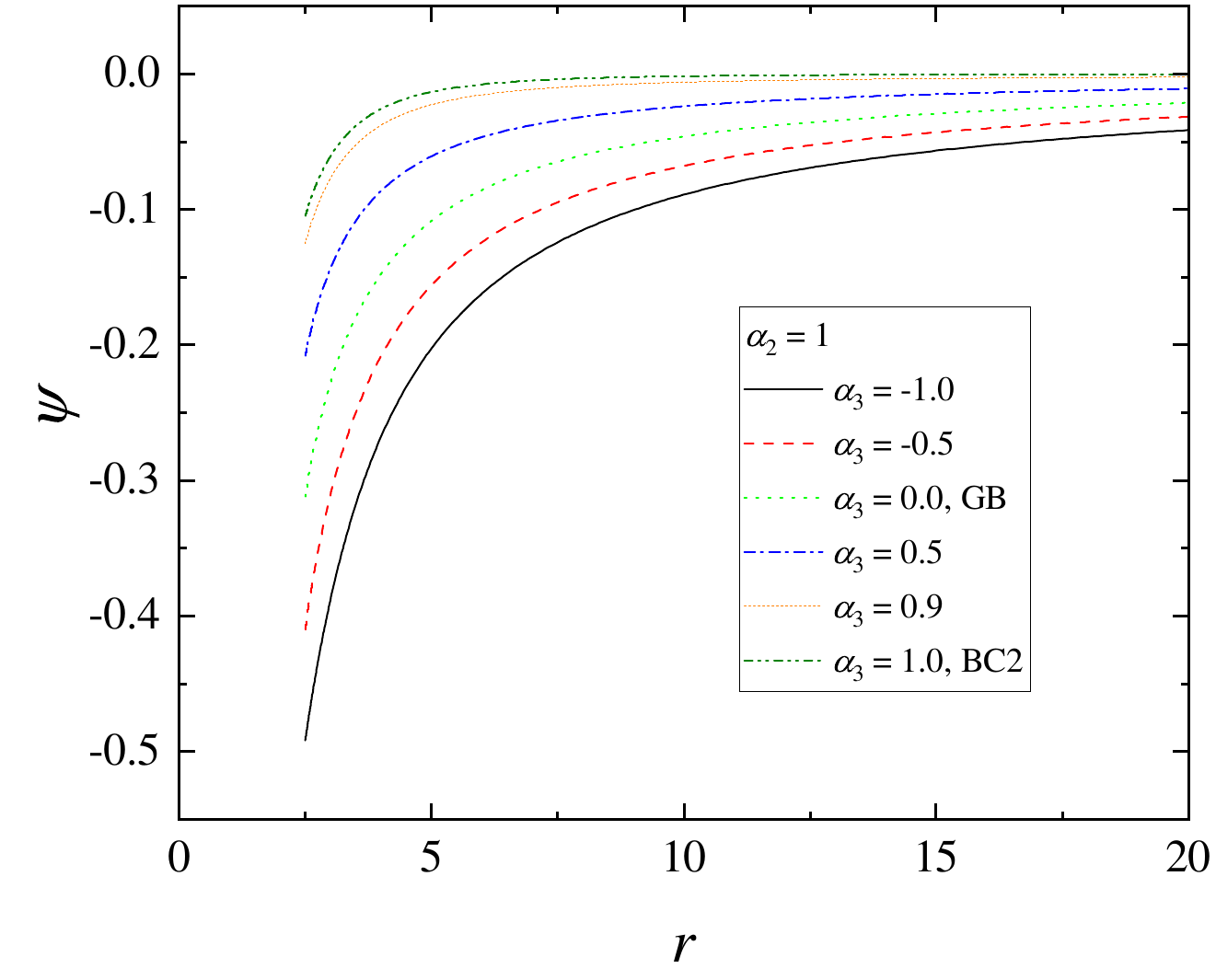}
    \includegraphics[scale=0.35]{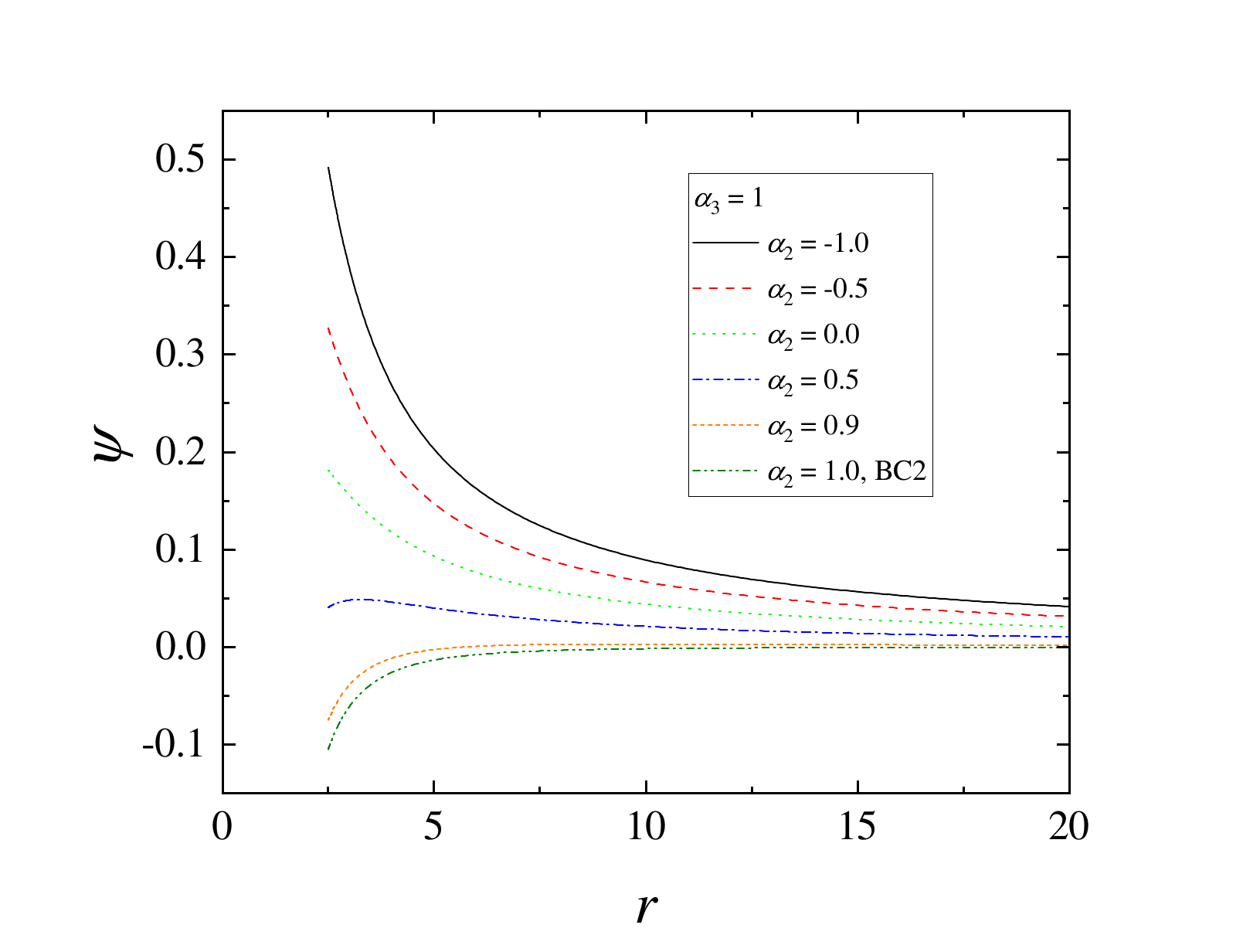}
    \caption{The radial profile of the scalar field for a linear coupling function \eqref{eq:LinCoupling}. All models are with a fixed $r_H=2.5$. We have worked either with $\alpha_2=1$ and varying $\alpha_3$ (\textit{left column}), or with $\alpha_3=1$ and varying $\alpha_2$ (\textit{right column}). }
    \label{fig:LinCoupling_scalar_field}
\end{figure}

A number of interesting observations can be made based on Fig. \ref{fig:LinCoupling}. As seen in the upper panels, in certain more extreme cases, there is a turning point of the branch, and the mass starts to increase with decreasing $r_H$. This is normally associated with an instability of the associated black hole solutions \cite{Blazquez-Salcedo:2018jnn,Silva:2018qhn}. For the plotted combinations of parameters, this is most evident in the top-right panel in Fig. \ref{fig:LinCoupling}; see the magenta line with $\alpha_2=-0.5$ and $\alpha_3=1$. The scalar charge $D$, on the other hand, depicted in the middle panels, is either positive or negative depending on the values and the signs of $\alpha_2$ and $\alpha_3$. As a consequence, there is a range of parameters where the scalar charge is practically vanishing. Interestingly, in Fig. \ref{fig:LinCoupling}, this happens for $\alpha_2=\alpha_3$ where BC2 is used. A combination of parameters that represent this interesting behavior is $\alpha_2=\alpha_3=1$, for which the BC2 should be used, depicted with dashed orange lines in Fig. \ref{fig:LinCoupling}. In such cases, the scalar field on the horizon is still large (see the bottom panels in the figure). This has interesting implications. For example, if put in a binary, such a black hole will emit only very little scalar dipole radiation while the scalar field might influence the binary dynamics significantly. 

To understand better the solutions in the shift symmetric theory, we present in Fig. \ref{fig:LinCoupling_scalar_field} the radial profiles of the scalar field. Contrary to the case of spontaneous scalarization considered in \cite{Bahamonde:2022chq} where very peculiar behavior was observed for small horizon radius black holes, here the scalar field is always monotonous. Perhaps this is due not only to the change of the coupling function but also to the fact that for this theory the solutions do not reach very close to $r_H=0$. They are instead terminated due to a violation of regularity on the horizon. 

Another interesting question is whether the purely teleparallel case $\alpha_2=0$ (pure $T_G$ coupling) or $\alpha_2=-\alpha_3$ (pure $B_G$ coupling) would differ from the Riemannian Gauss-Bonnet branches with ($\alpha_3=0$). Even though all these cases share a lot of qualitative similarities, some distinctions can be noted. For pure $T_G$ or $B_G$ coupling, the branches reach a maximum mass leading to the formation of a turning point as commented above. In addition, in these two cases, the deviations from GR seem to be stronger compared to the Riemannian Gauss-Bonnet theory.

\subsection{Pure quadratic coupling and spontaneous scalarization}\label{sec:numquad}
\begin{figure}[ht!]
    \includegraphics[scale=0.35]{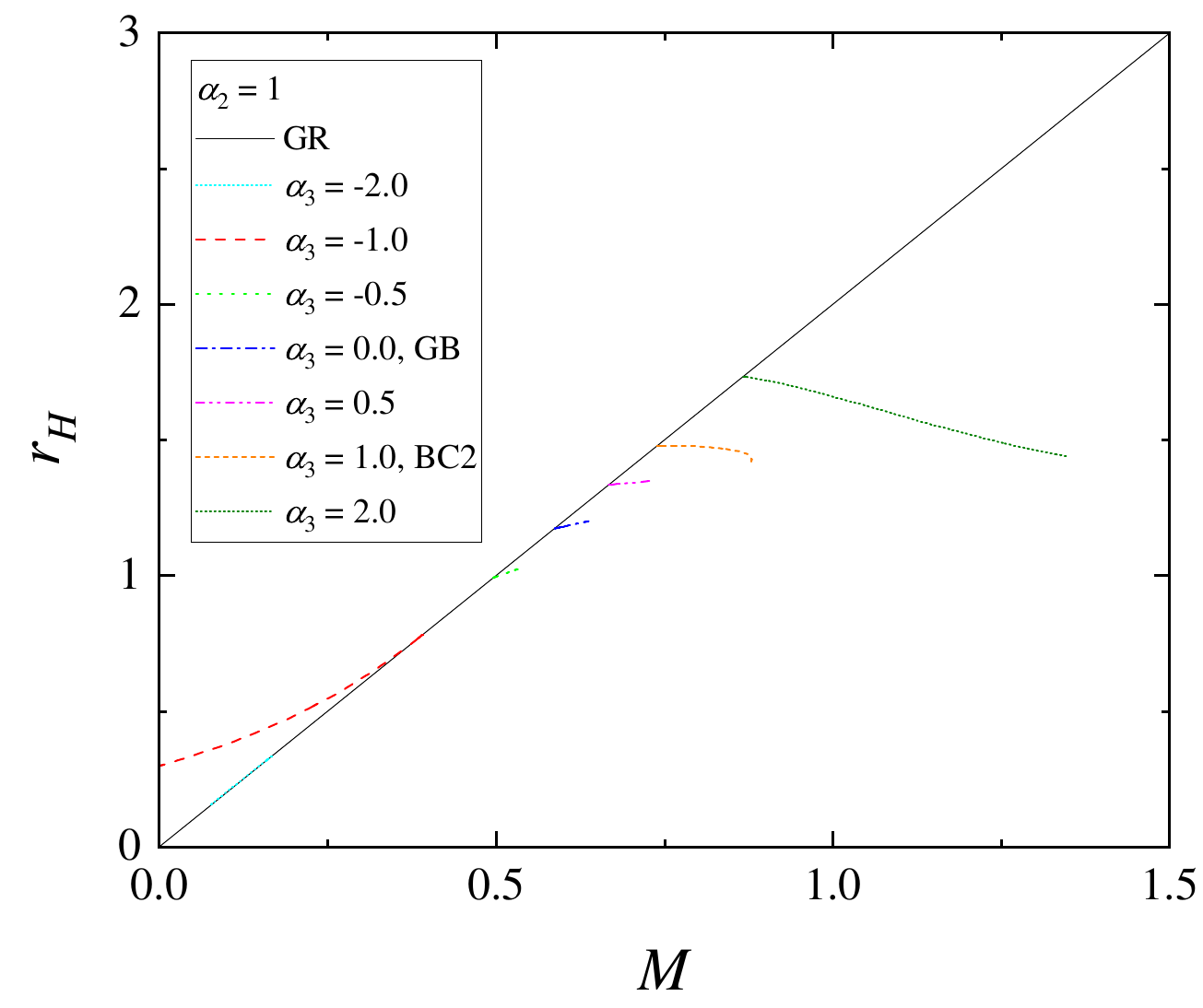}
    \includegraphics[scale=0.35]{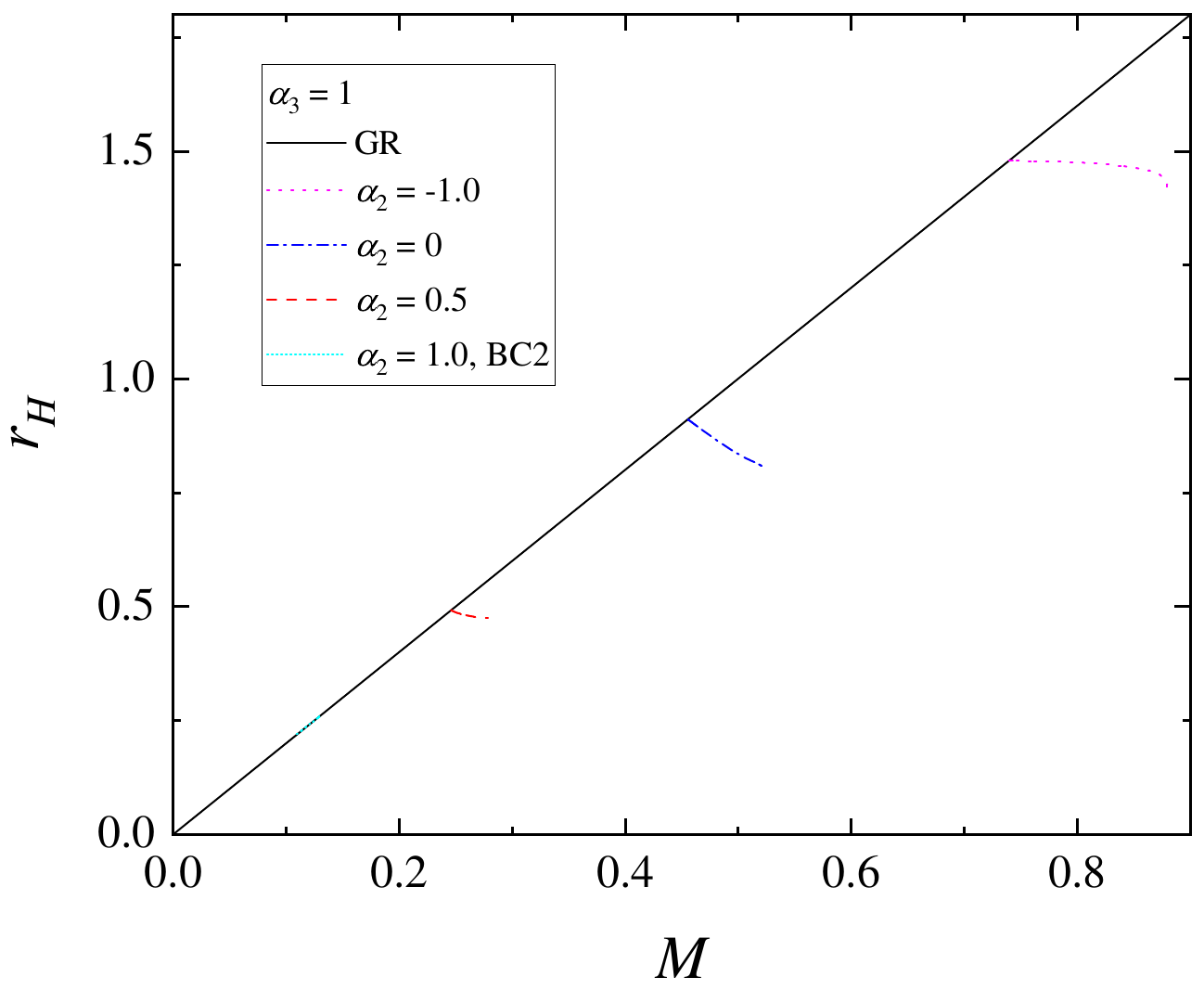}
    \includegraphics[scale=0.35]{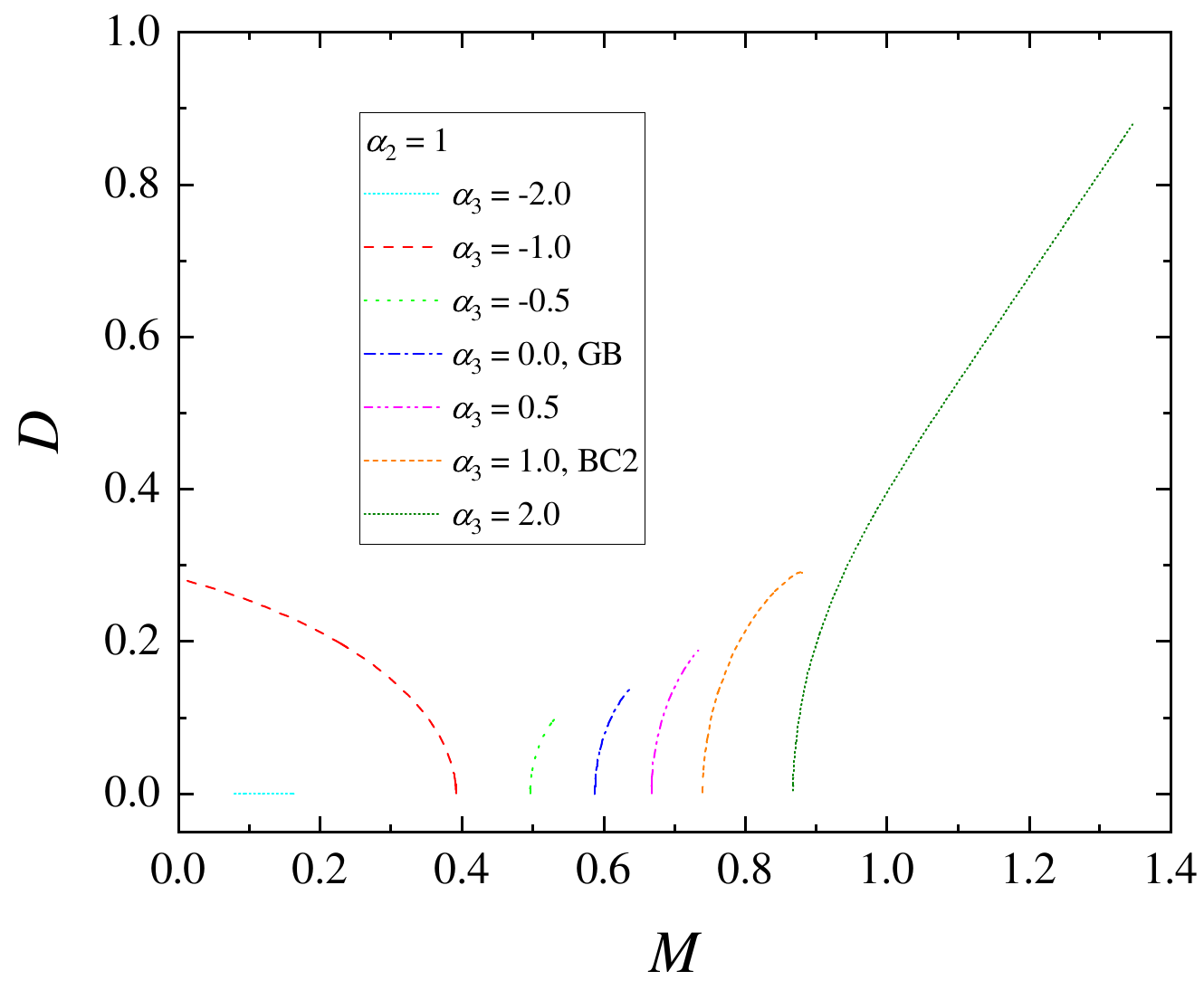}
    \includegraphics[scale=0.35]{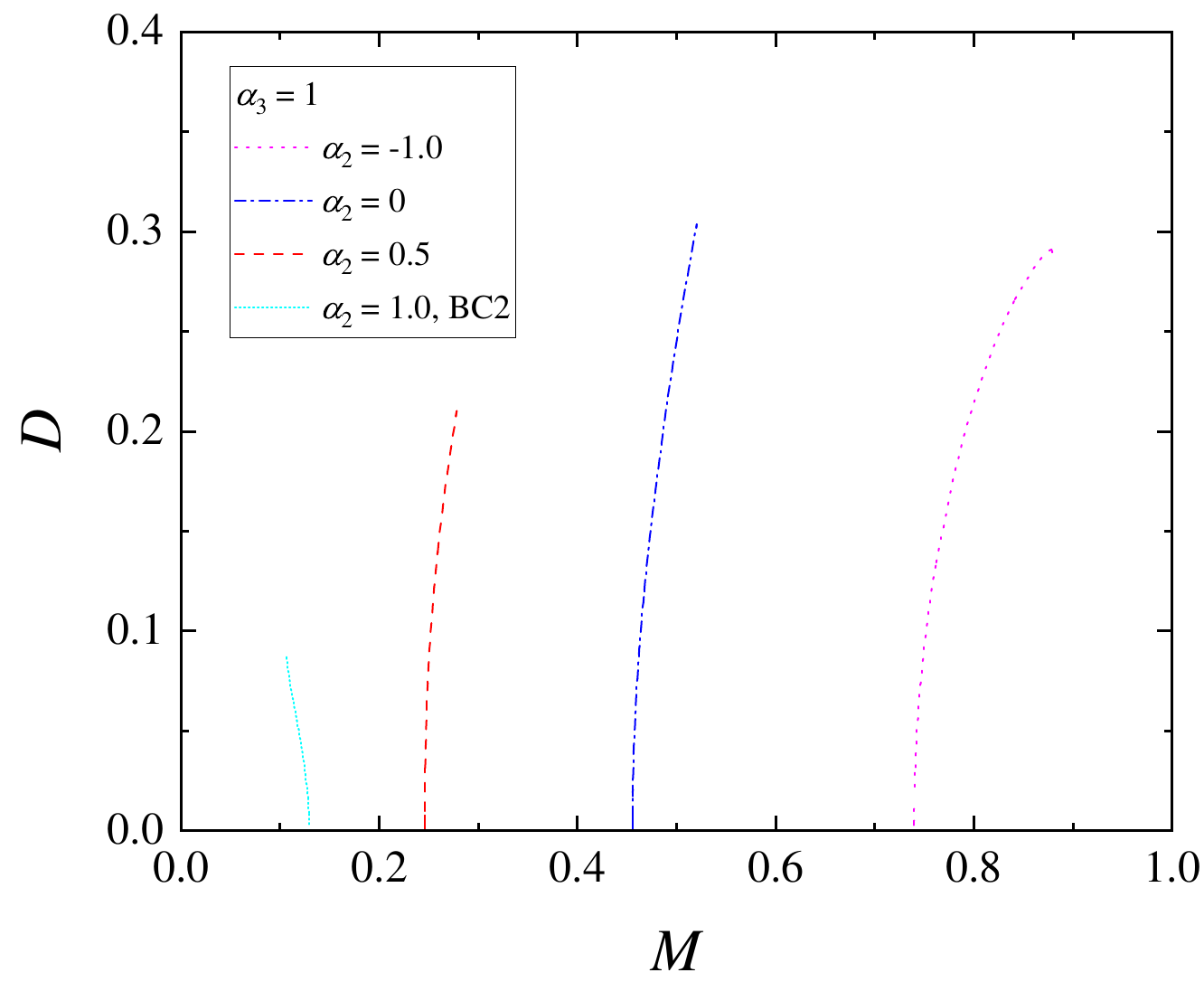}
    \includegraphics[scale=0.35]{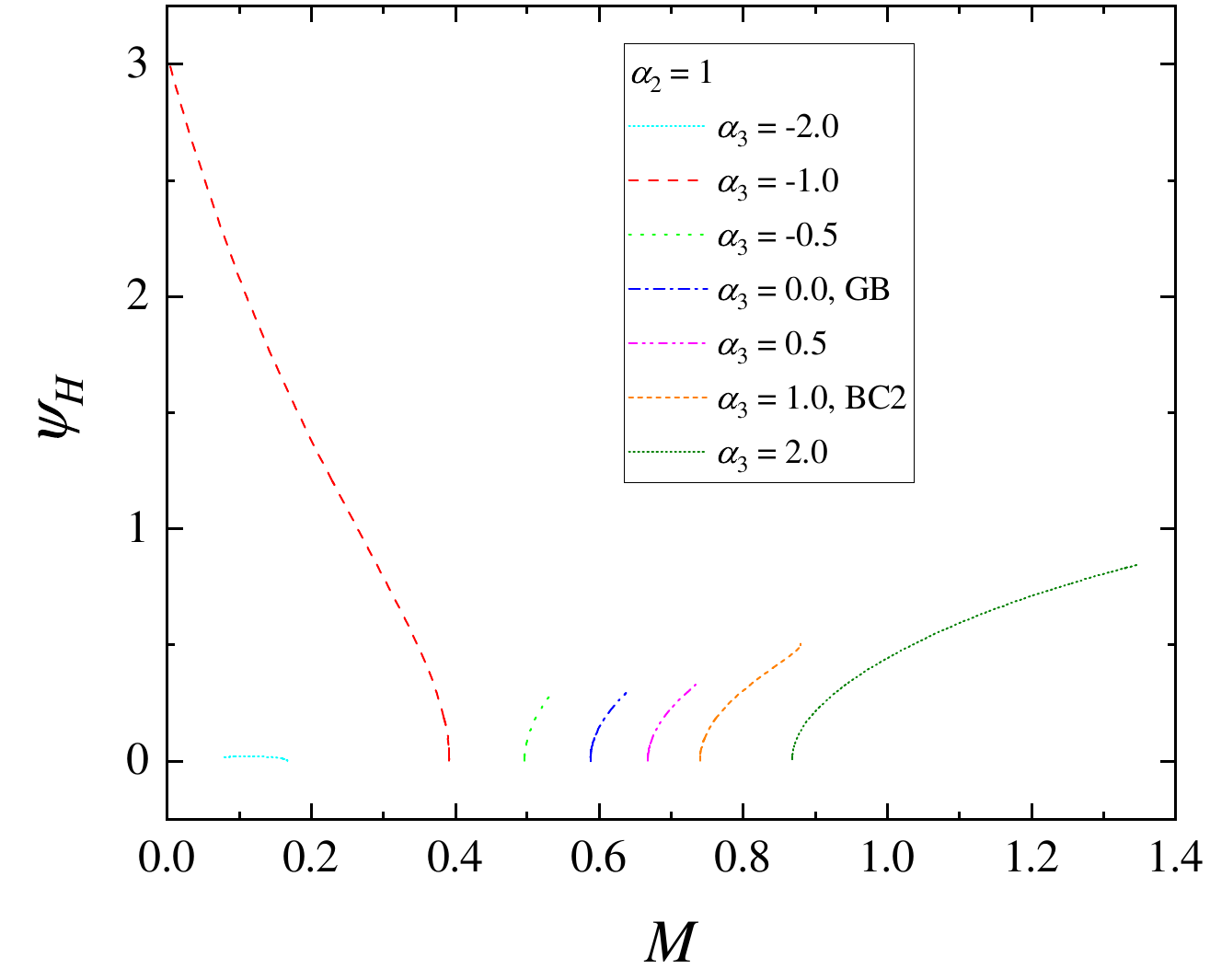}
    \includegraphics[scale=0.35]{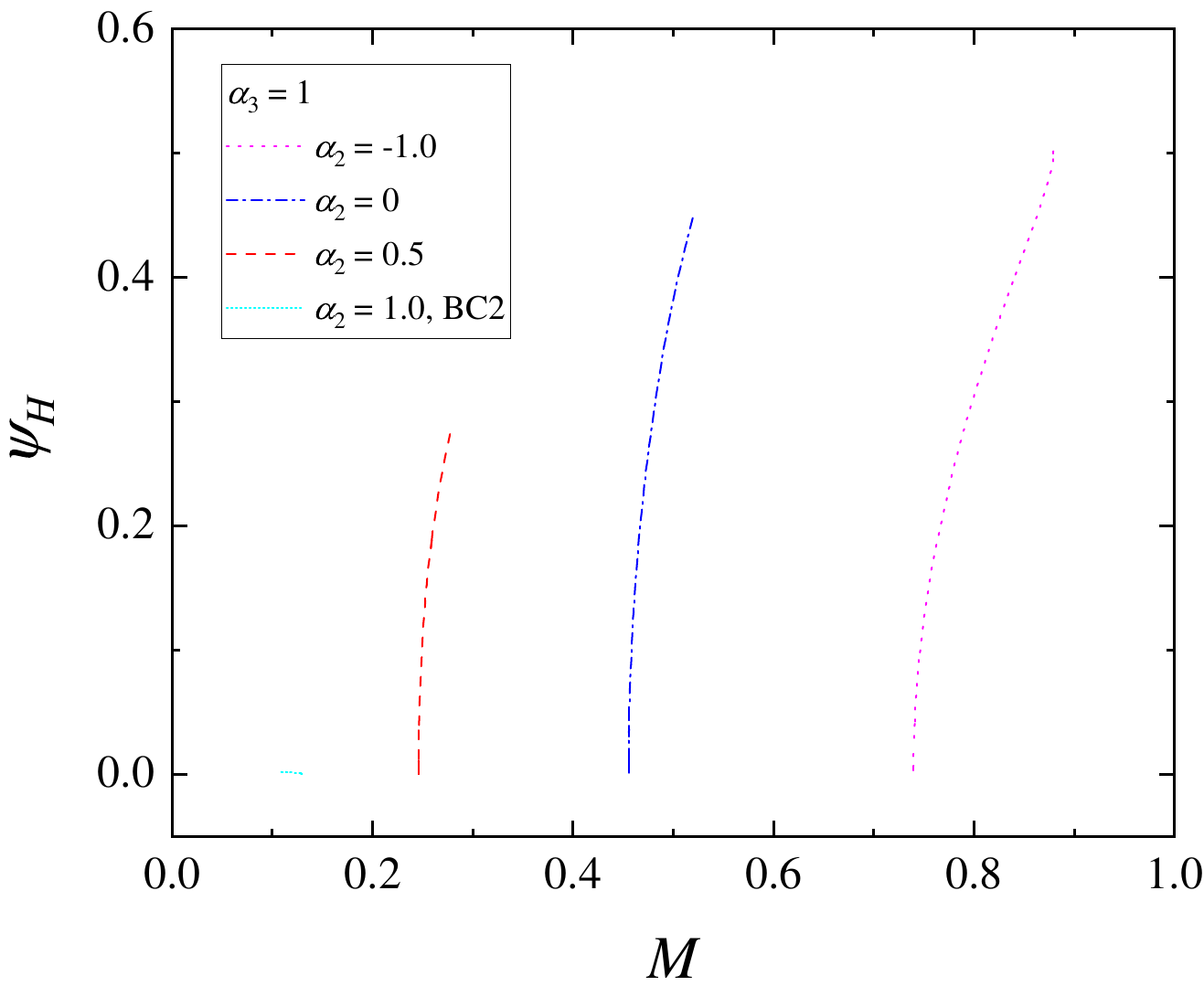}
    \caption{The radius of the horizon (\textit{top panels}), the scalar charge (\textit{middle panels}), and the scalar field at the horizon (\textit{bottom panels}) as functions of mass for a quadratic coupling function \eqref{eq:QuadraticCoupling}. The notations are similar to Fig. \ref{fig:LinCoupling}.}
    \label{fig:QuadraticCoupling}
\end{figure}

Let us now consider the case of spontaneous scalarization but with a different, simpler, coupling function compared to \cite{Bahamonde:2022chq}
\begin{align}\label{eq:QuadraticCoupling}
   \mathcal{G}_2=\mathcal{G}_3=\psi^2.
\end{align}
Such a coupling is often avoided in numerical simulations since, in the Riemannian case at least, it leads to unstable black hole solutions. In addition, the branches of solutions are terminated shortly after the bifurcation point. As we will show below, though, this behavior might change if we generalize these results to the teleparallel Gauss-Bonnet gravity.

In Fig. \ref{fig:QuadraticCoupling}, we plot the same characteristic as in the previous subsection, but for the coupling \eqref{eq:QuadraticCoupling}. Again, the two columns correspond to fixing one of the parameters, either $\alpha_2$ or $\alpha_3$, to unity and varying the other one. Note that in this case the GR-like black holes with zero scalar field are always solutions of the field equations and only in a certain range of black hole masses do additional scalarized solutions appear.

Here, again, we have combinations of parameters for which counterintuitive behavior is observed -- the black hole mass increases with the decrease of $r_H$. This is best evident in the green curve in Fig. \ref{fig:QuadraticCoupling}, top-left panel, and it is normally a sign of instability. However, all branches that turn right after the bifurcation point in the $D(M)$ diagram are most probably unstable if we borrow an analogy with the Riemannian Gauss-Bonnet case \cite{Blazquez-Salcedo:2018jnn,Silva:2018qhn}. The peculiar branches with decreasing $r_H$ when $M$ increases also turn right which is another sign of their instability.

Interestingly, for certain combinations of $\alpha_2$ and $\alpha_3$, predominantly when the pure teleparallel term is stronger, the behavior of the solutions changes -- the branch of solutions (in the middle and the lower panels of Fig. \ref{fig:QuadraticCoupling}) turn left after the bifurcation point instead of turning right. One example is the branch with $\alpha_2=1$ and $\alpha_3=-1$ that corresponds to pure $B_G$ coupling.  Such behavior is an indicator of stabilizing the solutions, at least in the  Riemannian Gauss-Bonnet case \cite{Silva:2018qhn}. Another point in favour of stability is that the horizon radius of the scalarized black holes, in this case, gets larger than the GR one contrary to all branches that turn right after bifurcation. Even though an expression for the black hole entropy in teleparallel gravity is not available yet and, in general, is very difficult to derive, this observation speaks in favour of the conjecture that the hairy black holes are thermodynamically preferred in that case. Thus, our results indicate that the pure teleparallel term might potentially lead to a stabilization of the black holes for pure quadratic coupling. A rigorous proof of (in)stability would require solving the linearized perturbation equations. Black hole perturbations and thermodynamics of black holes, though, are very involved in teleparallel gravity, and this is a problem that has not been properly addressed yet. Thus, we leave it for future work.

Let us again comment on the limiting cases of purely teleparallel theory with $\alpha_2=0$ (pure $T_G$ coupling) and $\alpha_2=-\alpha_3$ (pure $B_G$ coupling). In the former case, no qualitative deviation from the Riemannian theory was observed and only quantitative differences are present. Pure $B_G$ coupling, though, leads to the appearance of potentially stable branches as commented above. Interestingly, this happens only for one of the choices of signs of $\alpha_2$ and $\alpha_3$, namely   $\alpha_2=1,\; \alpha_3=-1$. The branch with opposite signs, $\alpha_2=-1,\; \alpha_3=1$, has the same qualitative behavior as in the pure Riemannian theory.

\section{Conclusions}\label{sec:concl}
In the present paper, we have examined in detail the existence and properties of black holes with scalar hair in teleparallel Gauss-Bonnet theory. In comparison to previous results where an exponential coupling function quadratic in the scalar field was considered \cite{Bahamonde:2022chq}, we have examined a shift-symmetric flavor of the theory with a linear coupling to the scalar field as well as a monomial quadratic coupling admitting spontaneous scalarization of black holes. In both cases, we have found very interesting qualitative behaviour, induced by torsion, different from the pure Riemannian theory.

We have performed the numerical calculations only in the case of complex tetrads \eqref{tetrad2}. We did so since, examining in detail the real tetrads, we have found strong evidence that in order to ensure regularity at the horizon, the field equations should reduce to the pure Riemannian case, at least to the lowest orders. Thus, we deem this case as not particularly interesting and only the complex tetrads can provide new phenomenology. The non-regularity of the horizon for the real tetrads has been also pointed out in~\cite{Awad:2022fhx} where it is argued that they might not be able to provide regular horizons. The argument is that it might be not possible for a real tetrad to smoothly provide the transition needed for a horizon, which is, changing from a spacelike vector field inside the horizon to a timelike one outside of it. Let us remark here that since the dynamical variable of our theory is the tetrads fields, an important study for the future is to investigate the conditions and constraints one needs to impose on dynamical complex tetrads, so that all possible observables stay real. In this article, all the observables we have derived are indeed real quantities. However, it is essential to explore whether this statement holds true in a broader context, warranting further investigation.

The first coupling function we have considered is proportional to $\psi$, that is the teleparallel analog to the shift-symmetric Riemannian Gauss-Bonnet gravity. A new interesting feature is that, for a certain parameter range, one can observe branches of black hole solutions with vanishing scalar charge but nonzero scalar field close to the black hole horizon. Such compact objects would not emit scalar gravitational radiation, despite their sometimes large deviation from GR. Even though one needs a bit of fine-tuning of the parameters in order to achieve such configurations, this is nevertheless an interesting theoretical observation. Another interesting distinct property is that some of the branches with a stronger contribution of the pure teleparallel term might possess a turning point close to their minimum mass which is a signal of instability.

The second coupling function we have considered is proportional to the square of the scalar field. Such a choice is the simplest one that can produce spontaneously scalarized black holes. In the pure Riemannian case, it always leads to unstable solutions. For a strong enough teleparallel contribution and proper signs of the coupling constants, we have discovered that some of the branches show strong signs of stability similar, e.g., to Refs.~\cite{Minamitsuji:2018xde,Macedo:2019sem}. Namely, after the bifurcation point, the branches turn left instead of right. In addition, contrary to the other combinations of parameters, they have larger horizon radius compared to GR, which is a sign that they might be thermodynamically favourable. A rigorous proof of stability, though, will of course require more sophisticated methods such as linear perturbations analysis or a derivation of the black hole entropy. Both topics are much more involved compared to pure GR and they have not yet been addressed. Thus, we leave them for future studies. 

The effects presented in this paper that distinguish the teleparallel, torsion induced, GB gravity from its Riemannian counterpart are more ``global'' and concern the qualitative behavior of the branches of solutions. On the other hand, the spontaneous scalarization with exponential coupling function in teleparallel GB gravity has shown some very interesting peculiarities in the solution profiles close to the zero black hole mass limit \cite{Bahamonde:2022chq}. The combination of both solidifies the conjecture that the teleparallel terms can lead to very interesting qualitative changes and the intuition built from the Riemannian Gauss-Bonnet theory, or other theories admitting black hole scalarization, cannot be easily transferred to the teleparallel case. Thus, it would be interesting to consider the dynamics of these objects and search for distinct astrophysical signatures. Moreover, the next step in the program of studying torsion-induced scalarization is going beyond vacuum solutions and studying, for example, the properties of neutron stars with scalar hair, which are not at all or have been very rarely studied in teleparallel gravity so far \cite{Ulhoa:2012pw,Solanki:2021fzo,Vilhena:2023srq}.

That being said, we should acknowledge that the study of the dynamical evolution of compact objects in teleparallel GB gravity (or many other teleparallel theories) is technically involved and will be investigated during future research projects. Indeed, already in the pure (Riemannian) Horndeski case, the study of dynamical evolution, e.g. for a binary black hole or neutron star systems, is still a subtle topic in general, see \cite{Ripley:2022cdh} for a review or \cite{Shibata:2022gec} for a recent study in a specific example. Performing similar studies in the teleparallel case would require a careful analysis of the well-posedness of those problems in general and the development of new techniques and algorithms able to tackle the specificities of non-Riemannian theories of gravity.

Another intriguing avenue of exploration involves delving into the symmetric teleparallel formulation of the scalar Gauss-Bonnet gravity theory, which solely relies on non-metricity~\cite{Armaleo:2023rhj}. Within these theories, the Gauss-Bonnet invariant can also be divided into two distinct terms. Consequently, it presents an engaging opportunity to seek out scalarized black hole solutions within this framework and subsequently conduct a comparative analysis with respect to the torsional case.

\section*{Acknowledgements}
This study is in part financed by the European Union-NextGenerationEU, through the National Recovery and Resilience Plan of the Republic of Bulgaria, project No. BG-RRP-2.004-0008-C01. SB is supported by JSPS Postdoctoral Fellowships for Research in Japan and KAKENHI Grant-in-Aid for Scientific Research No. JP21F21789.  DD acknowledges financial support via an Emmy Noether Research Group funded by the German Research Foundation (DFG) under grant no. DO 1771/1-1.  L.D. is supported by a grant from the Transilvania Fellowship Program for Postdoctoral Research/Young Researchers (September 2022).  CP is funded by the excellence cluster QuantumFrontiers funded by the Deutsche Forschungsgemeinschaft (DFG, German Research Foundation) under Germany’s Excellence Strategy – EXC-2123 QuantumFrontiers – 390837967. SB, LD and CP would like to acknowledge networking support by the COST Action CA18108.
	
\bibliographystyle{utphys}
\bibliography{references}

\end{document}